\documentclass[final]{pasj02}

\Received{$\langle$reception date$\rangle$}
\Accepted{$\langle$acception date$\rangle$}
\Published{$\langle$publication date$\rangle$}
\newcommand{\secref}[1]{section~\ref{#1}}
\newcommand{\equaref}[1]{equation~(\ref{eq:#1})}
\newcommand\figref[1]{{figure~\ref{fig:#1}}}
\newcommand\Figref[1]{{Figure~\ref{fig:#1}}}

\newcommand{\appref}[1]{Appendix~\ref{#1}}
\newcommand{\tabref}[1]{table~\ref{table:#1}}
\begin{document}

\title{The Star Formation and Chemical Evolution Histories of Ursa Minor Dwarf Spheroidal Galaxy}
\author{
{Kyosuke S. \textsc{Sato}\orcid{0000-0001-8239-4549},\altaffilmark{1,2}$^{*}$}
Yutaka \textsc{Komiyama},\altaffilmark{3}
Sakurako \textsc{Okamoto},\altaffilmark{1,2,4}
Masafumi \textsc{Yagi},\altaffilmark{2,3}
Itsuki \textsc{Ogami},\altaffilmark{1,2,5}
Mikito \textsc{Tanaka},\altaffilmark{3}
Nobuo \textsc{Arimoto},\altaffilmark{2}
Masashi \textsc{Chiba},\altaffilmark{6}
Evan N. \textsc{Kirby},\altaffilmark{7} and
Rosemary F.G. \textsc{Wyse},\altaffilmark{8}
Rintaro \textsc{Mori}\altaffilmark{3}}
\altaffiltext{1}{Astronomical Science Program, The Graduate University for Advanced Studies, SOKENDAI, 2-21-1 Osawa, Mitaka, Tokyo 181-8588, Japan}
\altaffiltext{2}{National Astronomical Observatory of Japan, 2-21-1 Osawa, Mitaka, Tokyo 181-8588, Japan}
\altaffiltext{3}{Department of Advanced Sciences, Faculty of Science and Engineering, Hosei University, 3-7-2 Kajino-cho, Koganei, Tokyo 184-8584, Japan}
\altaffiltext{4}{Subaru Telescope, National Astronomical Observatory of Japan, 650 North A'ohoku Place, Hilo, HI 96720, USA}
\altaffiltext{5}{The Institute of Statistical Mathematics, 10-3 Midoricho, Tachikawa, Tokyo 190-8562, Japan}
\altaffiltext{6}{Astronomical Institute, Tohoku University, Aoba-ku, Sendai, Miyagi 980-8578, Japan}
\altaffiltext{7}{Department of Physics and Astronomy, University of Notre Dame, Notre Dame, IN 46556, USA}
\altaffiltext{8}{Department of Physics and Astronomy, Johns Hopkins University, Baltimore, MD 21218, USA}

\email{kyosuke.sato@grad.nao.ac.jp}

\KeyWords{Local Group --- galaxies: dwarf --- galaxies: star formation --- galaxies: evolution --- galaxies: photometry}

\maketitle

\begin{abstract}
	We derive the star formation history (SFH) and chemical evolution history (CEH) of the Ursa Minor (UMi) dwarf spheroidal galaxy (dSph). We detect two distinct stellar populations that exist over 6 times half-light radius from its center.  
	The results are obtained by applying a newly developed algorithm to the deep and wide-field photometric dataset taken with Hyper Suprime-Cam on the Subaru Telescope.  
	The algorithm employs the genetic algorithm and the simulated annealing to minimize a $\chi^{2}$ value between the observed color-magnitude diagram (CMD) and synthetic CMD generated from the stellar isochrones.
	The age and metallicity resolutions are set to 0.5 Gyr and 0.1 dex, respectively.
	The accuracy assessment with mock galaxies shows that it returns the peaks of metallicity distributions and star formation period within 1 $\sigma$ of input value in the case of a single population.  
	In tests with two populations, two distinct metallicity peaks are identified without an offset from the input values, indicating the robustness of this algorithm.	
	The detected two populations in the UMi dSph have the metallicity peaks of [Fe/H] = $-2.2$ and $-2.5$; the metal-rich population started its star formation about 1 Gyr later than the metal-poor one.  
	The SFH of both metal-rich and metal-poor populations varies with distance from the center of the UMi dSph, without any age-gradients.
	These results suggest that the UMi dSph underwent a complex formation process, contrary to the simple formation history of dwarf galaxies previously thought.
\end{abstract}

\section{Introduction}
	In the $\Lambda$CDM model, structures in the Universe, such as galaxies, have been formed in the density fluctuations of dark matter \citep{1988ApJ...327..507F}.  
	The density fluctuation is enhanced with time, and the cosmic structure grows from a smaller scale (such as dwarf galaxies) to a larger scale (large galaxies and galaxy clusters) following the initial power-spectrum of density fluctuations in the $\Lambda$CDM model. 
	In this context, dwarf galaxies are the smallest and first formed systems, i.e., the smallest building blocks in the hierarchical clustering process, and some of them are considered as the survivors of the first galaxies. 
	Since the chemical composition of stars in dwarf galaxies contains information on their first billion years, they are a key to investigate galaxy formation process in the early Universe \citep{2015ARA&A..53..631F}, especially in the so-called epoch of re-ionization in the Universe \citep{2022NatAs...6...48A}.
	
	The morphology, gas content, and star formation history (SFH) of dwarf galaxies in the Milky Way (MW) vary with their distance from the center of the MW \citep{1999IAUS..192...17G}.
	With the exception of the Large Magellanic Cloud and the Small Magellanic Cloud, dwarf irregular galaxies (dIrrs) that still contain gas and maintain star formation are primarily located beyond 300 kpc from the center of the MW \citep{2014ApJ...795L...5S}.
	These dIrrs are only weakly affected by the  tidal forces of the MW.
	%Dwarf Irregular Galaxies (dIrr) which still contain gas and maintain star formation activity are mainly distributed beyond 300 kpc from the MW (except LMC, and SMC) and are weakly affected by tidal force from the MW. 
	Dwarf spheroidal galaxies (dSphs) that have already lost gas and quenched star formation are located within $\sim300$ kpc from the center of the MW and are affected by its strong environmental perturbations (e.g., \cite{2021ApJ...913...53P} and references therein).

	Based on their star formation histories, Galactic dSphs are classified into two types known as fast and slow dwarfs (e.g., \cite{2015ApJ...811L..18G}). 
	In this paper, we refer to those classified as dSphs among fast and slow dwarf as fast dSph and slow dSph, respectively.
	The fast dSph (e.g., Ursa Minor, Draco, Sculptor, Sextans, Cetus, and Tucana) quenched star formation in the early epoch.
	Therefore, about 90\% of their stellar component is formed during the first starburst phase.
	Slow dSphs, such as Fornax, Leo I, Leo A, and Carina, are characterized by extended SFHs, having formed only a small fraction of their stellar mass at early cosmic times and continuing to form stars until relatively recently. 
	These galaxies host an intermediate-age stellar population, reflecting a prolonged and complex evolutionary history.
	These differences probably depend on the formation and evolution process of each dwarf galaxy \citep{1987ApJ...322L..59S}.
	According to \citet{2015ApJ...811L..18G}, different SFHs between fast and slow dSphs could depend on the early environmental conditions rather than the result of a current morphological transformation driven by environmental effects like tidal stripping. 
	\citet{2014ApJ...789..147W} suggests that variation of SFH links to the first infall timing and environmental processing effects. 
	From the orbital analysis of MW satellites, quenching time of the fast dSphs (e.g., Sextans, Sculptor, Ursa Minor, Carina) coincide with the first infall, whereas for the ultra-faint dwarfs (UFDs; $L<10^5 L_\odot$), the quenched time does not coincide with the first infall (e.g., \cite{2019MNRAS.483.4031R}, \cite{2019arXiv190604180F}, \cite{2020ApJ...905..109M}).
	%These findings suggest that the variety of SFHs of Galactic dSphs are depended on a combination of early environmental conditions, infall times, and environmental effects.
	%The Galactic dSphs, that appear to be simple systems composed of old stars, could be more complex than previously considered simple stellar population \citep{WYSE2002395}.
	These findings suggest that Galactic dSphs could be more complex than previously thought as a simple system composed of a simple stellar population.
	They could have undergone a unique formation and evolutionary process to reach its current property, by a combination of early environmental conditions, infall times, and environmental effects.
	%Each dSph could have undergone a unique formation and evolutionary process to reach its current state.
	%The different SFH between fast and slow dwarfs could depend on the density of the intial surrounding environment \citep{2015ApJ...811L..18G} and the current luminosity \citep{2014ApJ...789..147W}.
	%These slow dwarfs (e.g. Fornax, Leo I, and Carina) located in more outer region than the fast one.
	%Fast dwarfs are predicted to evolved in high-density environment, then fully effected by the stellar feedbacks and reionization (\cite{2015ApJ...811L..18G}.
	%Also, the small mass dSphs are thought to host the Simple stellar population (e.g. \cite{2007ApJ...670..313S}, \cite{2008ApJ...685L..43K}, and \cite{2015ARA&A..53..631F}).
	%According to \cite{2020ApJ...905..109M}, Galactic UFDs which has smaller stellar mass than classical dSphs are finished star formation activity before the first infall to Milky Way

	%Recent studies have found the substructures in Galactic dSphs. 
	%For example, the ring-like structure and satellites of dwarf galaxies were found in Sextans dSph \citep{2018MNRAS.480..251C}, and the stellar halo-like structure was found in nine dwarfs including UMi dSph \citep{2024MNRAS.527.4209J}. 
	Recent studies have found that the multiple stellar populations differ in the metallicity, spatial distribution, and velocity dispersions in several galactic dSphs; Ursa Minor (\cite{2020MNRAS.495.3022P}), Fornax (\cite{2006A&A...459..423B}, \cite{2011ApJ...742...20W}, \cite{2015MNRAS.454.3996D}, \cite{2022A&A...659A.119K}), Sextans \citep{2001MNRAS.327L..15B,2025arXiv250402787T}, Carina (\cite{2016MNRAS.457.1299K}, \cite{2016ApJ...830..126F}, \cite{2018MNRAS.481..250H}), and Sculptor (\cite{1999ApJ...520L..33M}, \cite{2004ApJ...617L.119T}, \cite{2008ApJ...681L..13B}, \cite{2011ApJ...742...20W}, \cite{2025A&A...699A.347A}).
	The metal-poor population tends to be more widely distributed and kinematically hotter than the metal-rich population.
	The multiple stellar populations in some of the slow dSphs (e.g., Fornax and Carina) could be attributed to a history of multiple star formation events.
	However, the fast dSphs that host multiple stellar populations (e.g., UMi dSph, Sextans dSph, and Sculptor dSph) do not show episodic star formation events like the slow dSphs. 
	Instead, they exhibit a short, intense burst of star formation before $z \sim 2$, and the presence of multiple stellar populations suggests a complex formation history for the fast dSphs in the early Universe \citep{2018MNRAS.480.1587S, 2019A&A...630A.116S}.
	The star formation may have been episodic, but on timescales that are unresolved with current data.
%	The multiple stellar populations have also been discovered in fast dSphs.
	% The observational evidence of the multiple stellar populations in fast dSphs suggests a formation scenario in which multiple populations form simultaneously. 
%	Thus, to form the multiple stellar population in the fast dSphs, formation scenarios, which form the multiple population at almost same time are needed.
	%Oppositely, the distinct metallicity between metal-rich and metal-poor population of other fast dSphs which correspond to Sculptor and Sextans are $0.8\ \rm{dex}$ (metal-rich:$\rm{[Fe/H]}\sim-1.5$ and metal-poor:$\rm{[Fe/H]}\sim-2.3$) and \citep{1999ApJ...520L..33M} $0.5\ \rm{dex}$ (metal-rich:$\rm{[Fe/H]}\sim-1.8$ and metal-poor:$\rm{[Fe/H]}\leq-2.3$) for Sextans \citep{2001MNRAS.327L..15B}, respectively.

	The UMi dSph is considered a candidate for a first galaxy, having formed its stars at very early Universe.
	This is supported by its low present-day stellar mass ($2.9 \times 10^{5}\ M_{\odot}$; \cite{2012AJ....144....4M}) and low mean metallicity ($[\mathrm{Fe/H}] = -2.13$; \cite{2011ApJ...727...78K}).
	These characteristics suggest that it can be regarded as one of the smallest building blocks in the hierarchical structure formation scenario.
	% The UMi dSph is the candidate first galaxy, which plausibly formed its stars very early, because of its current low stellar mass ($2.9\times10^{5}\ M_{\odot}$; \cite{2012AJ....144....4M}) and low mean metallicity ($[\rm{Fe/H}]=-2.13$; \cite{2011ApJ...727...78K}). 
	It is located at 69$\pm$4 kpc from the Sun \citep{1999AJ....118..366M}.
	Among the multiple stellar populations of fast dSphs, those in the UMi dSph are distinctive because the metallicities of the two populations are very close ($\sim0.25\ \rm{dex}$ between metal-rich: $\rm{[Fe/H]}=-2.05\pm0.03$ and metal-poor: $\rm{[Fe/H]}=-2.29^{+0.05}_{-0.06}$; \cite{2020MNRAS.495.3022P}).
	In the case of other fast dSphs, the metallicity difference between two populations is larger; $0.8\ \rm{dex}$ in Sculptor dSph ($\rm{[Fe/H]}\sim-1.5$ and $\rm{[Fe/H]}\sim-2.3$; \cite{1999ApJ...520L..33M}), $0.5\ \rm{dex}$ in Sextans dSph ($\rm{[Fe/H]}\sim-1.8$ and $\rm{[Fe/H]}\leq-2.3$; \cite{2001MNRAS.327L..15B}).
	This characteristic feature of the UMi dSph, having a fairly close metallicity compared to those of Sextans and Sculptor dSphs, could be attributed to the merger of two similarly sized galaxies, or little gas enrichment before the formation of the second stellar component \citep{2020MNRAS.495.3022P}.
	The SFH of the UMi dSph was inferred from the image taken by the 2.5 m Isaac Newton Telescope \citep{2002AJ....123.3199C} and the deep observational data obtained by the Hubble Space Telescope (HST; \cite{2014ApJ...789..147W}). 
%	In the case of \citep{2014ApJ...789..147W}, due to the small coverage of HST data, and not enough to search the spatial dependence of the star formation history.
	However, the data in \citet{2002AJ....123.3199C} are too shallow to resolve the age-metallicity degeneracy \citep{1999ASPC..192..283W}.
	Also, the data show a large scatter in the Main Sequence Turn-Off (MSTO), which has an important role to solve age-metallicity degeneracy. %\citep{1999ASPC..192..283W}.
	The HST studies \citep{2014ApJ...789..147W} cover a small spatial area, making insufficient to examine the spatial variation in the SFH.
%	For \citep{2002AJ....123.3199C}, the data are shallow that the MSTO observations needed to resolve the age-metallicity degeneracy \citep{1999ASPC..192..283W} are insufficient.

	Algorithms for solving the SFH and CEH of galaxies by linear combination of simple stellar populations (SSPs) have been studied by several groups (e.g., \cite{DOLPHIN1997397}, \cite{1997AJ....114..680A}, \cite{1999AJ....118.2262H}).
	The SFH and CEH have been derived for many dwarf galaxies within the Local Group using these techniques (e.g., \cite{2010ApJ...720.1225M, 2011ApJ...730...14H, 2012A&A...544A..73D, 2018MNRAS.480.1587S, 2019MNRAS.487.5862B, 2021MNRAS.508..245M, 2021MNRAS.501.3962R, 2021MNRAS.502..642R, 2022MNRAS.513L..40M, 2024ApJ...975...42C, 2024ApJ...976...60M}).
	% (e.g. \cite{2014ApJ...789..147W}, \cite{2011ApJ...730...14H}, \cite{2015ApJ...811...76C}).
	\citet{2013A&A...551A.103D} combines a photometric data with a spectroscopically derived metallicity distribution and age-metallicity relation of Fornax dSph.
	However, the resolution of the metallicities in these studies are not enough to divide the multiple stellar population of Galactic dSphs.
	A resolution of 0.1 dex for $[\rm{Fe/H}] <-2.0$ should be employed to estimate an SFH and CEH to separate the close multiple population of $\sim0.3$ dex of the UMi dSph.
	
	This study investigates the SFH of the UMi dSph resolved by metallicities and its spatial dependence using the Hyper Suprime-Cam (HSC), which has a very large field of view of 1.5 degrees on the 8.2 m Subaru Telescope.
	We apply a newly developed algorithm that simultaneously estimate the SFH and CEH with high resolution, 0.5 Gyr in age and 0.1 dex in metallicity, through synthetic CMD fitting. 
	This enables, for the first time, the derivation of spatially resolved SFH and CEH across the entire extent of UMi. 
	Our results confirm the presence of two distinct metallicity populations, in agreement with \citet{2020MNRAS.495.3022P}.	
	% This study investigates the SFH of the UMi dSph resolved by metallicities and its spatial dependence using the Hyper Suprime-Cam (HSC), which has a very large field of view of 1.5 degrees on the 8.2m Subaru Telescope.
	% We detect two distinct metallicity populations at [Fe/H] = $-2.2$ and $-2.5$.
	% While the SFHs of the metal-rich and metal-poor populations in the UMi dSph vary with the distance from the center of the UMi dSph, no clear age-gradients are observed.
	% The metal-rich population shows a 1 Gyr delay in the onset of star formation relative to the metal-poor population.
	This paper is organized as follows. 
	In \secref{data}, the details of data reduction and data qualities are summarized.
	In \secref{three}, the structural parameters of the UMi dSph are estimated.
	In \secref{amrd},  the methods, accuracy, and main results are described.
	In \secref{discussion} the color distribution analysis and the spacial variation of SFH are discussed.
	Finally, the summary of this work is given in \secref{summary}.
\section{Data and reduction}\label{data}
	\subsection{HSC observation}
\begin{figure}[h!]
					\begin{center}
					\includegraphics[width=9cm]{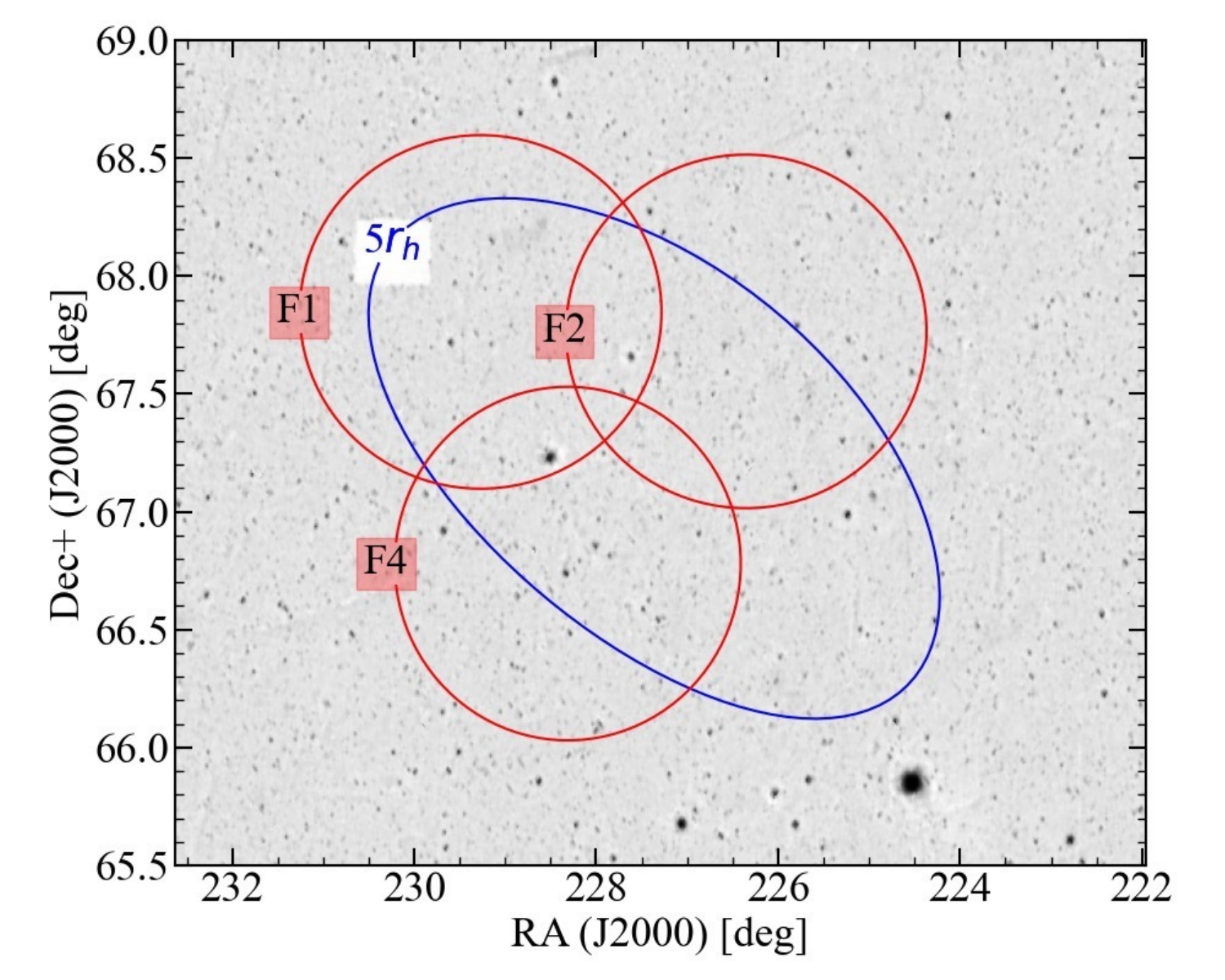}
					\end{center}
					\caption
					{
					Observed region around the UMi dSph. 
					Red circles correspond to our HSC pointings.
					%The exposure time and seeing information of each UMi\_F1, UMi\_F2, and UMi\_F3 are exhibited on the \tabref{observation}.
					The blue ellipse shows the 5 times half-right radius, $r_{\rm{h}}$, estimated in \secref{three}.
					The background is a PS1 $g$-band image.
					{Alt text: A plot showing astronomical image with right ascension on the x-axis and declination on the y-axis.}
					}\label{fig:FOV} 
				\end{figure}

			The photometric observation of the UMi dSph was conducted on the HSC mounted with the 8.2 m Subaru Telescope (Subaru/HSC) on 2015 May 25 (UTC) with $g$- and $i$-bands (proposal ID: S15A-OT08).
			The positions of observed fields are shown in \figref{FOV} as red circles overlaid on the $g$-band image of Pan-STARRS DR1 (PS1).
			In this observation, $g$- and $i$-band images were acquired under good seeing condition ranging from \timeform{0.43"} to \timeform{0.72"}.
			The details of observation are summarized in \tabref{observation}.
			%For all fields, we obtained a total exposure time of 1980 s, 1800 s, 1800 s, and 3600 s, 3780 s, 3960 s in $g$- and $i$-bands (UMi\_F1, UMi\_F2, and UMi\_F3), respectively.
			The excellent photometric conditions and sufficient exposure times allow us to reach down to MSTO at the distance of the UMi dSph.
			MSTO stars are essential for estimating star formation history and metallicity while resolving the age-metallicity degeneracy on the Red Giant Branch (RGB) stars \citep{1999ASPC..192..283W}.
				\begin{table*}[h*]
						\caption{Observation log.}
						\label{table:observation}%info of obs
					\centering
						\footnotesize
						\renewcommand{\arraystretch}{1.5} % 行間を広げる
						\begin{tabular}{cllllllll}
							\hline
							Filter & Field & R.A. [deg] &Dec. [deg] & Date & Seeing & Exposure time \\
							\hline
							$g$ & $\rm{UMi\_F1}$ & $229.26999$ & $67.8476$ & $2015\ \rm{May}\ 25$ & $\timeform{0.54"}$--$\timeform{0.72"}$ & $180\ \rm{s} \times11$\\
							$g$ & $\rm{UMi\_F2}$ & $226.34948$ & $67.7651$ & $2015\ \rm{May}\ 25$ & $\timeform{0.48"}$--$\timeform{0.66"}$ & $180\ \rm{s} \times10$\\
							$g$ & $\rm{UMi\_F4}$ & $228.31209$ & $66.7801$ & $2015\ \rm{May}\ 25$ & $\timeform{0.48"}$--$\timeform{0.63"}$ & $170\ \rm{s} \times10$\\
							$i$ & $\rm{UMi\_F1}$ & $229.26999$ & $67.8476$ & $2015\ \rm{May}\ 25$ & $\timeform{0.43"}$--$\timeform{0.52"}$ & $180\ \rm{s} \times21$\\
							$i$ & $\rm{UMi\_F2}$ & $226.34948$ & $67.7651$ & $2015\ \rm{May}\ 25$ & $\timeform{0.43"}$--$\timeform{0.54"}$ & $180\ \rm{s} \times20$\\
							$i$	& $\rm{UMi\_F4}$ & $228.31209$ & $66.7801$ & $2015\ \rm{May}\ 25$ & $\timeform{0.48"}$--$\timeform{0.66"}$ & $180\ \rm{s} \times22$\\
							\hline
						\end{tabular}
				\end{table*}
%%%%%%%%%%%%%%%%%%%%%%%%%%
	\subsection{Reduction}\label{reduction}
		We reduce this data using the HSC pipeline version 8.5.3 ($\rm{hscPipe}$; \cite{2018PASJ...70S...5B}). 
		$\rm{hscPipe}$ was developed by the National Astronomical Observatory of Japan (NAOJ), Princeton University, and Kavli Institute for the Physics and Mathematics of the Universe (Kavli IPMU). 
		$\rm{hscPipe}$ is based on the pipeline of Vera C. Rubin Observatory (VRO; \cite{2008AIPC.1082..359I, 2017ASPC..512..279J}) and it is optimized for the HSC. 
		$\rm{hscPipe}$, firstly, carries out bias subtraction, dark subtraction, flat fielding, and sky subtraction for each CCD frame.
		Secondly, the photometric zero point estimation and astrometry are performed by comparing bright objects with the Pan-STARRS1 catalog (\cite{2012ApJ...750...99T, 2016arXiv161205560C}) for all frames \citep{2018PASJ...70S...6H}.
		Then, all frames are co-added into a large image based on the mosaicking method in each band.
		The cosmic rays are removed in this operation.
		From the co-added images of $g-$ and $i-$bands, Kron, CModel, and the point-spread function(PSF) magnitudes are derived	\citep{2018PASJ...70S...5B}.
		In $\rm{hscPipe}$, PSFEx \citep{2011ASPC..442..435B, 2013ascl.soft01001B} is used to estimate the PSF.	
		$\rm{hscPipe}$ provides the $\rm{Extendedness}$ parameter, which judges whether the detected objects are point sources or not, and it is based on a difference between CModel and PSF magnitudes.
		If the difference is smaller than 0.0164 mag, the object is judged as a point-source \citep{2018PASJ...70S...5B}.
		We simply use this $\rm{Extendedness}$ parameter to select stellar objects from the photometric catalog.

		The Galactic extinctions are corrected using the Galactic dust map by \citet{2011ApJ...737..103S}.
		The $E(B-V)$ values for all objects across our observed fields are taken from the NASA/IPAC Infrared Science Archive\footnote{https://ned.ipac.caltech.edu}.
		For estimation of extinction value $\rm{A(\lambda)}$, we derive the effective wavelength of the model star when observed by Subaru/HSC $g$- and $i$-bands filter systems.
		We use the spectral energy distribution of an F5V-type star, which has an effective temperature of ${T_{\rm{eff}}}=6440\ \rm{K}$ and surface gravity of $\log {g}=4.34$, from the Castelli \& Kurucz 2004 Stellar Atmosphere Models\footnote{https://www.stsci.edu/hst/instrumentation/reference-data-for-calibration-and-tools/astronomical-catalogs/castelli-and-kurucz-atlas} \citep{2003IAUS..210P.A20C} to estimate the effective wavelength. 
		In addition, to reproduce the observed spectrum by Subaru/HSC, we multiply by the system throughput (i.e., atmospheric extinction, throughput of optics, transmittance of filters, and quantum efficiency of CCD\footnote{https://www.subarutelescope.org/Observing/Instruments/HSC/index.html}; \cite{2018PASJ...70...66K, 2018PASJ...70S...1M}).
		Effective wavelengths are derived as 4904.15 \AA\ and 7786.93 \AA\ for $g$- and $i$-bands, respectively. 
		We use the extinction curve, which shows the relationship between Galactic dust reddening and extinction value at a given wavelength \citep{1999PASP..111...63F}, and then correct the extinction value for every star according to its position.
		The extinction-corrected magnitudes for each $g$- and $i$-band are 
			\begin{eqnarray}
				g_{0} &=& g-3.102\times E(B-V) \notag \\
				i_{0} &=& i-1.583\times E(B-V). 
				\label{extinction}
			\end{eqnarray}
		The subscript 0 to the magnitude or the color means those after the extinction correction (or the value not affected by the Galactic extinction) throughout this work.
		The mean extinction values in the $g$- and $i$-bands are 0.09 mag and 0.04 mag, respectively.
		
\begin{figure}[h!]
				\begin{center}
				\includegraphics[width=8cm]{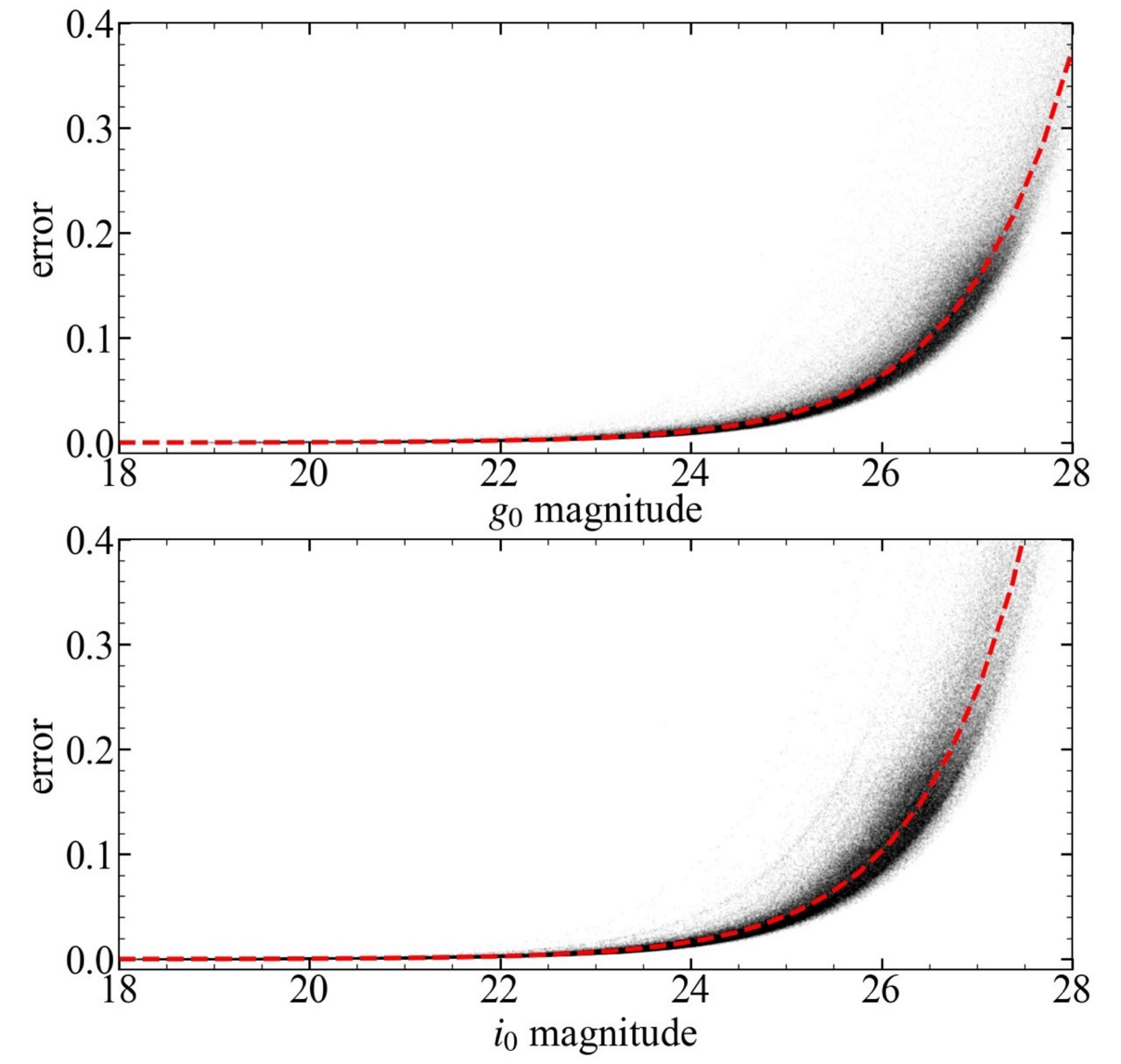}
				\end{center}
				\caption{
				The relationship between magnitude and photometric error of stars in the UMi dSph catalog in both $g$-band (top) and $i$-band (bottom).
				Red dashed lines indicate the result of the exponential fitting.
				{Alt text: A plot with magnitude on the x-axis and photometric error on the y-axis, showing a trend of increasing error at fainter magnitudes.}
				\label{fig:photerr} }%labelはcaption内に書く必要あり
			\end{figure}

		The photometric error of the point-sources given by hscPipe against their magnitudes are shown in \figref{photerr}.
		We conduct 2 $\sigma$ clipping three times to fit the photometric error functions against both magnitudes.
		We conduct the exponential curve fitting to each of the $g$- and $i$-bands.
			
\begin{figure}[h]
				\begin{center}
				\includegraphics[width=8.8cm]{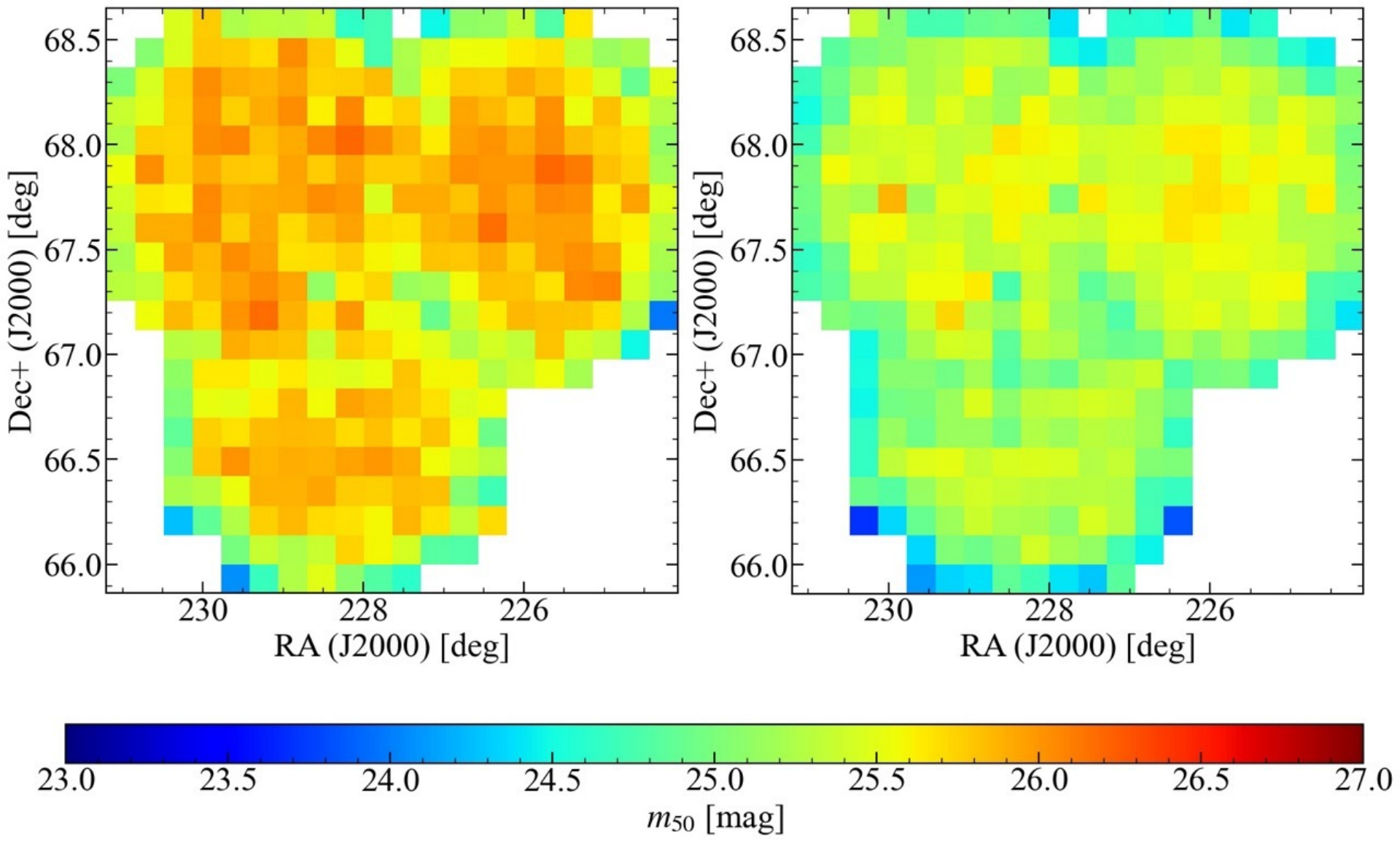}
				\end{center}
				\caption{
				The $g$- and $i$-bands 50\% completeness magnitude maps.
				The left panel is for $g$-band, and the right panel is for $i$-band.
				%In this artificial stars test, we divide the whole data into $20\times20$ areas.
				The color correspond to the 50\% detection completeness magnitude, which derived by fitting the theoretical model indicated in \equaref{comp_fit} to the observed data.
				{Alt text: Two two-dimensional histograms.}
			\label{fig:compmap} }
			\end{figure}

		To derive the detection completeness of photometric data, we run the artificial star test using the same method as in \citet{2025MNRAS.536..530O}.
		Artificial stars are created using the PSF models, and injected into the original images at interval of $60\times60\ \mathrm{pixel^2}$.
		These stars are added to the original images in magnitude intervals of 1.0 mag, ranging from 16.0 mag to 28.0 mag in both the $g$- and $i$-bands.
		We then apply the same source detection and photometric routines of $\rm{hscPipe}$ to the images containing the artificial stars.
		During this process, an artificial star is considered detected 
		if the detected position is within \timeform{0.5"} of the injected coordinate and the detected magnitude is within 0.5 mag difference from the input magnitude.
		%Otherwise, it is classified as non-detected. 
		We fit the result by the completeness function model of \citet{2016ApJ...833..167M},
			\begin{equation}
				\label{eq:comp_fit}
					\eta(m)=\frac{A}{1+\exp{(\frac{m-m_{50}}{\rho}})},
			\end{equation}
		where $m$ is the apparent magnitude of the $g$- and $i$-bands, $m_{50}$ is the apparent magnitude of 50\% completeness, and $A$ and $\rho$ are the model parameters.
		These parameters, $m_{50}$, $A$ and $\rho$ are estimated by the least-squares fitting.
		\Figref{compmap} displays the $m_{50}$ maps of the $g$- and $i$-band photometric data, with the entire observed field divided into $20 \times 20$ regions. 
		The maps indicate that the $m_{50}$ values in both bands are fainter, i.e., indicating higher completeness, at the centers of the HSC pointings.
		The detection completeness in the CMD is evaluated for each of the $20 \times 20$ regions shown in \Figref{compmap}, and the resulting completeness values are used to assign weights to individual stars.
%%%%%
\section{Photometric properties of the Ursa Minor dSph}\label{three}
	\subsection{Member selection with Color magnitude diagram}
\begin{figure}[h]
					\begin{center}
					\includegraphics[width=9cm]{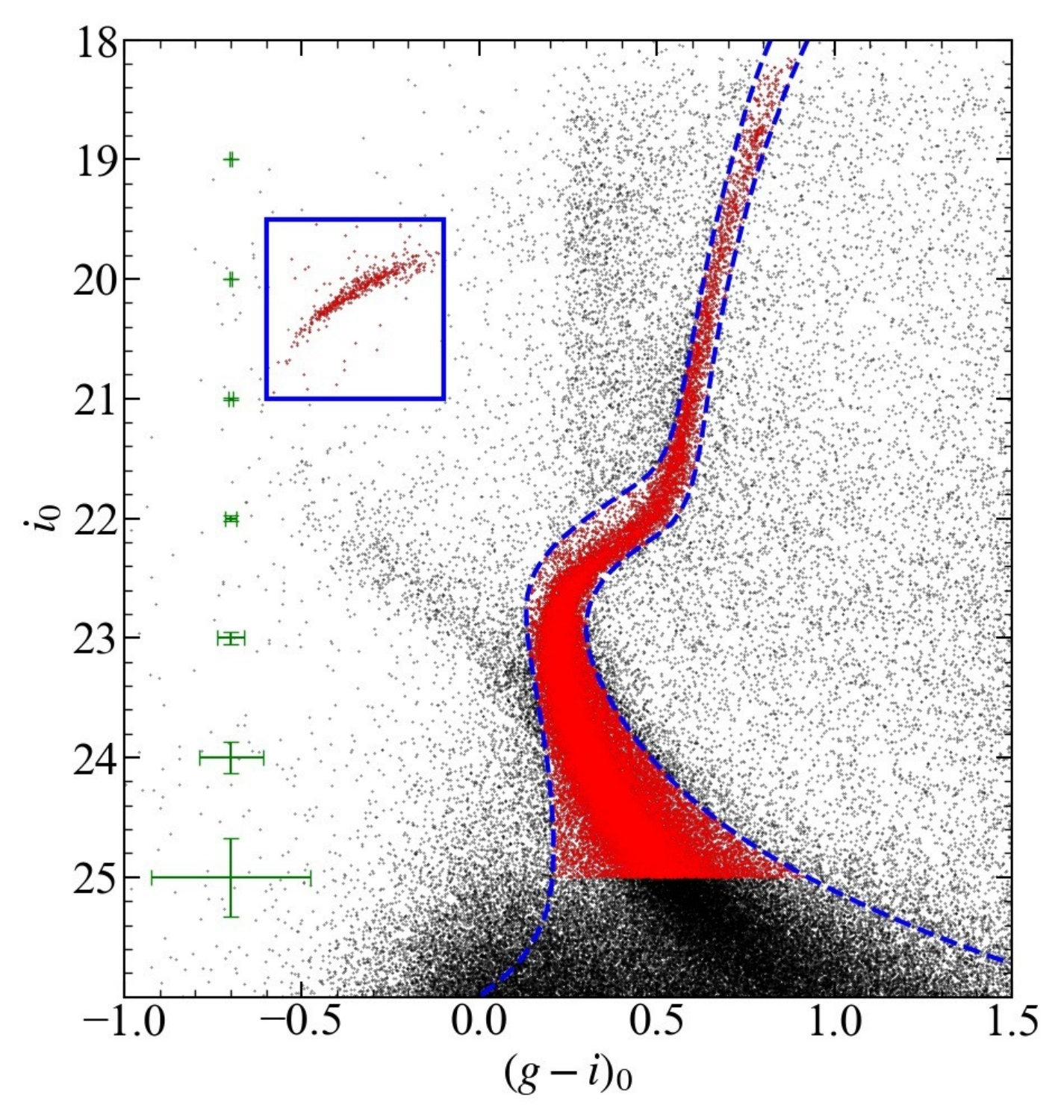}
					\end{center}
					\caption{
						The CMD of point-sources within the HSC footprint around the UMi dSph. 
						The red points indicate the objects deemed likely members of the UMi dSph, which is selected by the isochrone filter.
						This isochrone filter is indicated by blue dashed lines showing the boundary of the region on CMD, which is drawn in the range of $12\leq\rm{age}\leq\ 13\rm{[Gyr]}$ and $-3.2\leq\rm{[Fe/H]}\leq-1.8\ \rm{[dex]}$.
						The blue rectangle spanning $-0.6<(g-i)_{0}<-0.1, 19.5<i_{0}<21$ is set to select BHB stars.
						The green error bars show the $8\sigma$ photometric error used to make isochrone filter.
						{Alt text: A graph.}
				\label{fig:cmd_sel}}
				\end{figure}

			\Figref{cmd_sel} shows the CMD of point-sources within the entire HSC footprint in the UMi dSph. 
			The CMD shows the well-defined RGB to MS sequence with significant foreground and background contamination.
			%The CMD is extremely high quality, almost unaffected by photometric errors up to MSTO.
			Various evolutionary stages of the UMi dSph stars are observed in this plot. 
			RGB sequence is seen at around $i_{0} \leq 22$ mag and $0.5 <(g-i)_0 < 1.0$, which is useful for estimating the stellar metallicity. 
			The sub-giant branch (SGB) and main sequence stars at $22 \leq i_{0} \leq 23$ are good tracers of stellar age. 
			These RGB, SGB, and MSTO distributions suggest the UMi dSph is composed of old and metal-poor ($\rm{[Fe/H]}\leq-2.0$) stellar populations.
			Blue horizontal branch (BHB) stars are located around $i_{0} \sim 20$ mag and $(g-i)_0 \sim -0.3$.
			BHB stars are also old and metal-poor stellar populations.
			The apparent absence of red horizontal branch stars  indicates that the UMi dSph is dominated by metal-poor stars.
			The foreground contamination from the MW dwarf stars is located at $(g-i)_{0} \gtrsim 0.3$.
			Background galaxies are distributed around $(g-i)_{0} \sim 0.25$ and $i_{0}\sim 25.5$ mag \citep{2011ApJS..195...18R}.
			The blue diagonal sequence located around $22\lesssim i_{0}\lesssim23$ and $-0.5\lesssim (g-i)_{0} \lesssim0.1$ is the blue straggler stars \citep{2007MNRAS.380.1127M}.

			To determine the structural properties of the UMi dSph, we perform the member selection on the CMD with the isochrone filter.
			We make the isochrone filter using the BaSTI isochrone (\cite{2018ApJ...856..125H}, \cite{2021ApJ...908..102P}) in the range of $12\leq\rm{age}\leq\ 13\ \rm{Gyr}$ and $-3.2\leq\rm{[Fe/H]}\leq-1.8\ \rm{dex}$ with combining the $8\sigma$ of the photometric error value of each $g-$ and $i-$band estimated in \figref{photerr}.
			The $8\sigma$ of photometric error range is set by eye to cover the broad distribution of MS stars, including binaries.
			This roughly selected range does not significantly affect the results, as the stellar density model used for the estimation of structural parameters includes the background profile.
			We cut the faint objects at $i_{0}\geq25$, since the background galaxies are prominent in that range.
			In addition, we set the rectangle to select the BHB stars.
			An object is deemed as a member star of the UMi dSph if it satisfies either of the aforementioned criteria.
			The total number of selected member stars based on these criteria is 48,887.
	\subsection{Structural properties}\label{structuralProperties}
			The structural properties of the UMi dSph are estimated using the iterative Bayesian approach of Markov Chain Monte Carlo (MCMC) analysis.
			The MCMC method is used to perform fitting based on the log-likelihood described in \equaref{likelihood} for the spatial distribution of the UMi dSph.
			The log-likelihood is described as
				\begin{equation}
					\label{eq:likelihood}
						P_{\rm{tot}}=\sum_{k}\log\left[\frac{\rho_{\rm{model}}({d_{k}}|\mathcal{P})}{\int_{S}\rho_{\rm{model}}{\rm d}S}\right],
				\end{equation}
			where $\{{d_{k}}\}_{1\leq k\leq N_{\rm{tot}}}$ is the coordinate datasets $\{x_{k},y_{k}\}$, $x_{k}$ and $y_{k}$ are the coordinates of each star in the projected sky coordinate. 
			When these equatorial coordinates of stars are converted to projected sky coordinates, we set the initial the UMi dSph center as $(\rm{R.A.}_0,\ \rm{Dec.}_0)=(227.2925,\ 67.21444)\ \rm{deg}$ (J2000) from \citet{2003ApJ...588L..21K}.
			The distance modulus is fixed to the literature value, $(m-M)_{0} = 19.18\pm0.12$ \citep{1999AJ....118..366M} as listed in \tabref{post}.
			$\mathcal{P}$ is a parameter set of estimation, and $\rho_{\rm{model}}$ is the radial density profile including the member stars and contamination.
			$\rho_{\rm{model}}$ is normalized to the expected number of stars in the region $S$ \citep{2016ApJ...833..167M}.
			$\mathcal{P}$, the parameter set of $x_{0},\ y_{0},\ r_{\rm{h}},\ \epsilon,\ \theta$, and $\ N_{\ast}$, is estimated in this MCMC analysis. 
			The $x_{0}$ and $y_{0}$ are the center coordinate of the UMi dSph.
			The $\epsilon$ represents the ellipticity, which is defined by the major $(a)$ and minor $(b)$ axes as $\epsilon=1-(b/a)$, and $r_{\rm{h}}$ represents the half-light radius.
			In this study, we focus on the half-density profile of stellar counts, since we use the resolved stellar data of the UMi dSph. 
			Then, assuming a uniform stellar population that is not affected by mass segregation, this quantity is equivalent to the half-light radius.
			The $\theta$ is the position angle of the major axis.
			The position angle is defined such that $0^{\circ}$ points to the north and $+90^{\circ}$ to the east.
			The $N_{\ast}$ is the number of the UMi member stars.
			We use the $\text{EMCEE}$ module \citep{2013PASP..125..306F} for calculation.
	%	The priors are set as \tabref{prior}.
		%20240216
			%%prior_table
				\begin{table}[h]
					\caption{The priors used in $\rm{EMCEE}$.}
					\label{table:prior}%info of obs
				\centering
					\footnotesize
					\renewcommand{\arraystretch}{1.5} % 行間を広げる
					\begin{tabular*}{8cm}{@{\extracolsep{\fill}}cc}
						\hline
						MCMC prior parameter & Parameter range \\
						\hline
						$x_{0}\ [\rm{pc}]$ & $-1500\leq x_{0}\leq1500$\\
						$y_{0}\ [\rm{pc}]$ & $-1500\leq y_{0}\leq1500$\\
						$r_{\rm{h}}\ [\rm{pc}]$ & $0\leq r_{\rm{h}}\leq1500$\\
						$\epsilon\ $ & $0\leq\epsilon\leq1$\\
						$\theta\ [^{\circ}] $ & $0\leq\theta<180$\\
						$N_{\ast}$ & $0\leq N_{\ast}\leq N_{\rm{tot}}$\\
						\hline
					\end{tabular*}
				\end{table}\\
			For analysis, we use the radial density profile model of a dwarf galaxy (e.g., \cite{2016ApJ...833..167M, 2022MNRAS.509...16M}), $\rho_{\rm{dwarf}}(r)$, which is described as
				\begin{equation}
					\label{eq:rho_dwarf}
					\rho_{\rm{dwarf}}(r)=\frac{a^{2}}{2\pi r_{\rm{h}}^{2}(1-\epsilon)}N_{\ast}\exp(-ar/r_{\rm{h}}).
				\end{equation}
			The coefficient $a=1.67834...$ is obtained analytically by solving
			\begin{equation}
				\label{eq:rh_re}
				1-\exp(-a)(1+a)=\frac{1}{2}.
			\end{equation}
			The variable $r$ of \equaref{rho_dwarf} represents an elliptical radius, which is written as
				\begin{eqnarray}
					r &=& \bigg( \left\{ \frac{1}{1-\epsilon} \left[ (x - x_{0}) \cos\theta - (y - y_{0}) \sin\theta \right] \right\}^2 \nonumber \\
					&& \quad + \left[ (x - x_{0}) \sin\theta + (y - y_{0}) \cos\theta \right]^2 \bigg)^{\frac{1}{2}},
					\label{eq:elliptical}
				\end{eqnarray}
			where $x$ and $y$ are the coordinates of each star in the projected sky.
			The background surface density profile is described as
				\begin{equation}
					\label{eq:bg_con}
					\Sigma_{b}=\frac{N_{\rm{tot}}-\int_{S}\rho_{\rm{dwarf}}{\rm d}S}{S},
				\end{equation}
			where $N_{\rm{tot}}$ is the number of stars selected by the CMD selection (see \figref{cmd_sel}).
			The area of the data, denoted by $S$, is calculated by assuming a circular radius of 2.0 kpc ($\sim\timeform{100.4'}$) and considering only objects within this circle for estimation.
			We also consider the background density profile \equaref{bg_con} as 
				\begin{equation}
					\label{eq:rho_model}
					\rho_{\rm{model}}=\rho_{\rm{dwarf}}(r)+\Sigma_{b}.
				\end{equation}
			The likelihood, given in \equaref{likelihood} used in MCMC analysis is obtained by combining equations (\ref{eq:rho_dwarf}), (\ref{eq:elliptical}), (\ref{eq:bg_con}), and (\ref{eq:rho_model}).

			In our MCMC analysis, we adopt the log-likelihood function defined in \equaref{likelihood} and assume uniform (flat) priors for all parameters to ensure full coverage of the parameter space (\tabref{prior}).
			In the $\rm{EMCEE}$ routine, we set 32 walkers, over a total of 8,000 iterations with a burn-in stage of 2,000.
			%The marginal posterior distribution, and estimated structure parameters are shown in \figref{corner} and summarized in \tabref{post}.
			We show the marginalized posterior distribution, the probability density function of the derived structure parameters in \appref{apd:cornerplot}.
			The derived parameters are summarized in \tabref{post}.
			Previously, \citet{2018ApJ...860...66M} estimated the ellipticity as $\epsilon=0.55\pm0.01$ based on limited sky coverage.
			Our estimated value $\epsilon=0.461\pm0.003$ is more circular than the result of \citet{2018ApJ...860...66M}.
			This distinct ellipticity depends on the size of the area used for fitting.
			According to \citet{2020MNRAS.495.3022P}, there are two populations with different ellipticities, metallicities and velocity dispersions.
			The extended distribution of the metal-poor population tends to be more circular than the metal-rich one.
			Therefore, the small ellipticity derived in this study could be influenced by these metal-poor populations than in previous studies because of the larger fitting region.
			This result is consistent with the metallicity gradient derived in \secref{AMRD_obs}.
				\begin{table}[h]
					\caption{The resulting structural parameters of the UMi dSph.}
					\label{table:post}
					\centering
					\footnotesize
					\renewcommand{\arraystretch}{1.5} % 行間を広げる
					\begin{tabular*}{9cm}{@{\extracolsep{\fill}}lll}
						\hline
						Property & Value & Reference\\
						\hline
						%$\rm{\alpha_{0}\ [^{\circ}]} $ (J2000)& $227.2420$ & \citet{2018ApJ...860...66M}\\
						%$\rm{\delta_{0}\ [^{\circ}]}$ (J2000)& $67.2221\pm0.001$ & \citet{2018ApJ...860...66M}\\
						%$r_{\rm{h}}\ [\rm{pc}]$ & $271\pm3$ & \citet{2018ApJ...860...66M}\\
						%$\epsilon\ $ & $0.55\pm0.01$ & \citet{2018ApJ...860...66M}\\
						%$\theta\ [^{\circ}]$ & $50\pm1$ & \citet{2018ApJ...860...66M}\\
						$(m-M)_{0}$ & $19.18\pm0.12$ & \citet{1999AJ....118..366M}\\
						$\rm{\alpha_{0}\ [^{\circ}]} $ (J2000)& $227.285\pm0.003$ & This work\\
						$\rm{\delta_{0}\ [^{\circ}]}$ (J2000)& $67.234\pm0.001$ & This work\\
						$r_{\rm{h}}\ [\rm{pc}]$ & $345.8^{+1.7}_{-1.6}$ & This work\\
						$r_{\rm{h}}\ [\rm{arcmin}]$ & $17.47^{+0.08}_{-0.08}$ & This work\\
						$\epsilon\ $ & $0.461\pm0.003$ & This work\\
						$\theta\ [^{\circ}] $ & $49.96\pm{0.26}$ & This work\\
						$N_{\ast}$ & $43725.1^{+78.2}_{-82.9}$ & This work\\
						\hline
					\end{tabular*}
				\end{table}				
	\subsection{Radial CMD morphology of the UMi dSph}
\begin{figure*}[h!]
					\begin{center}
					\includegraphics[width=18cm]{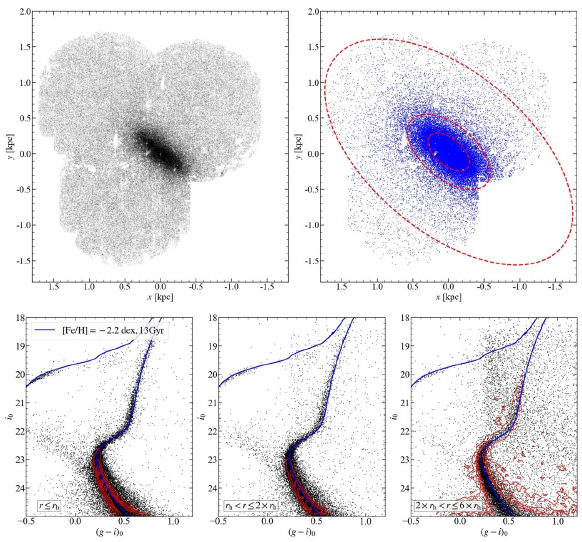}
					\end{center}
					\caption
					{
					The top left panel displays the spatial distribution of all the point sources.
					The top right panel shows the spatial map of the selected member stars by the isochrone filter and rectangle for BHB selection shown in \figref{cmd_sel}.
					Dashed red ellipses show $1$, $2$, and $6$ $r_{\rm{h}}$, respectively. 
					%The black and red cross indicate the center of projected sky coordinate and each annulus, respectively.
					The bottom three panels show the CMDs of selected member stars within $r \leq r_{\rm{h}}$, $r_{\rm{h}} < r \leq 2 r_{\rm{h}}$, and $2 r_{\rm{h}} < r \leq 6 r_{\rm{h}}$, respectively.
					Blue solid lines are isochrones from BaSTI with $\rm{[Fe/H]}=-2.2$, and $\rm{age}=13\ \rm{Gyr}$.
					%The red contours based on the $(g-i)_0\times i_0= 0.04 \times 0.14$
					Contours show number densities binned by 0.04 mag in the x-axis and 0.1 mag in the y-axis.
					The contours of left two diagrams are drawn every 100 from 200 to 700 stars per bin, and the right panel is every 25 from 10 to 110 stars per bin. 
					\label{fig:CMD} 
					{Alt text: Five scatter plots.}
					}
				\end{figure*}

			To investigate the difference between the inner and outer regions of CMD morphology, we divide the entire observed data into three annuli as shown in the top right panel of \figref{CMD}.
			These CMDs indicate that MSTO members in the UMi dSph are well detected, and the photometric errors of $g$- and $i$-bands are $<0.01$ at MSTO magnitude $\sim$23 mag. 
			This quality is enough to resolve the color degeneracy of age and metallicity as demonstrated in \secref{AMRD_obs}.
			The relative contamination by foreground stars increases in the outer region.
			The outermost CMD (bottom right panel of \figref{CMD}) corresponds to the region including the outer tidal radius, $\timeform{77.9'}$ \citep{2010MNRAS.406.1220W}.
			Nonetheless, we clearly confirm the existence of the MS, MSTO and a few RGB stars.
			In \secref{AMRD_obs}, we derive the SFH and CEH from these CMDs.
			The blue straggler stars are well confirmed in $r \leq 2 r_{\rm{h}}$.
			However, even taking into account the foreground contamination and the number of the UMi stars in the outer edge, the number ratio between blue straggler stars and horizontal branch stars in $r>2r_{\rm{h}}$ is smaller than in the inner regions.
			The detailed analysis of the MS stars will be presented by Sato et al. (in prep).
\section{Age-metallicity relation diagram}\label{amrd}
		The age-metallicity relation diagram (AMRD) is a 2-dimensional diagram of metallicity distribution of newly formed stars at each look-back time, which reveals the coevolution of star formation and chemical evolution of stellar systems.
		The configuration of AMRD reflects the difference in the formation scenario of a satellite \citep{2019MNRAS.488.2312G}.
		Therefore, we can use this information to understand how the UMi dSph has formed and evolved to what we see today.

		It is difficult to solve the SFH and the photometric metallicity distribution simultaneously, due to the color-degeneracy of stellar metallicity and age on the CMD.
		%However, the deep photometry covering bright RGB to faint old MS, such as our dataset taken by 8.2 m telescope Subaru/HSC, allows us to solve this degeneracy, since small photometric errors over quite a wide field of view with a diameter of 1.5 deg are achieved even for faint stars below the MSTO magnitude.
		%However, the deep photometry covering from the bright RGB to the faint old MS, such as our dataset taken with the 8.2 m Subaru Telescope and HSC, allows us to solve this degeneracy. 
		%This is because high photometric quality is achieved even for faint stars below the MSTO magnitude over a wide field of view with a diameter of 1.5 degrees.
		%The wide-field coverage of Subaru/HSC is well suited to study spatial characteristics of Galactic dwarf galaxies (e.g., radial population gradient from the center).
		However, the deep photometry covering from the bright RGB to the faint old MS, such as our dataset taken with the Subaru/HSC, allows us to solve this degeneracy. 
		This is because high photometric quality is achieved even for faint stars below the MSTO magnitude over a wide field of view with a diameter of 1.5 degrees. 
		Thanks to this wide-field, small photometric error and high detection completeness, we are able to derive the SFH and CEH out to regions beyond the tidal radius of the UMi dSph.
		The wide-area coverage of Subaru/HSC is thus particularly well suited for investigating the spatial characteristics of Galactic dwarf galaxies, such as radial gradients in stellar populations from the galaxy center.

		In sections \ref{IMF} to \ref{AMRD_obs}, we describe the procedure to derive the AMRD from our photometric catalog.
		In sections \ref{IMF} and \ref{SSP} we explain the procedure to make the SSPs.
		In \secref{HGA}, we explain how to estimate the AMRD from the observed CMD using the synthetic CMD using Hybrid Genetic algorithm (HGA).
		In \secref{test}, we assess the estimation accuracy of our algorithm using the mock galaxies.
		Finally, \secref{AMRD_obs} shows the resultant AMRDs of the UMi dSph.
		\subsection{The initial mass function}\label{IMF}
			In this study, we adopt the Kroupa initial mass function (IMF) \citep{2001MNRAS.322..231K, 2002ASPC..285...86K}, using the open-source Python library $\rm{imf}$\footnote{https://github.com/keflavich/imf} to make the mass distribution of stellar systems.
			This choice is motivated by previous studies of Ursa Minor, which indicate that its stellar population is dominated by old, metal-poor stars and shows no evidence for a non-standard IMF. 
			In particular, \citet{WYSE2002395} concluded that the IMF of the UMi dSph is consistent with that of the solar neighborhood.
			Therefore, the adoption of a Kroupa IMF in this study is well justified.
			In this Python library, the range of the mass function is $0.08\leq M/M_{\odot} \leq 100$, and the binary system is not considered.
			Since the flux of a binary system is the combined flux of the primary and companion stars, binaries tend to exhibit a broader distribution in the CMD compared to single stars.
			In our analysis, we adopt a binary fraction of $f_{\mathrm{bin}} = 0.5$, and assume a flat distribution for the mass ratio ($q = M_{\rm{companion}}/M_{\rm{primary}}$), where $q$ is uniformly distributed between 0 and 1.
			This choice is motivated by \citet{2018AJ....156..257S}, who constrained the binary fraction in Ursa Minor dSph to lie between $0.45^{+0.05}_{-0.05}$ and $0.96^{+0.03}_{-0.06}$ based on multi-epoch radial velocity data.
			\citet{2018AJ....156..257S} also reported that the mass ratios prefers a flat distribution.
			Our adopted parameters are within the range suggested by \citet{2018AJ....156..257S} and reflect currently plausible assumptions for binaries in the UMi dSph.

		\subsection{Simple stellar populations}\label{SSP}
			We generate 391 SSPs, which are pieces of the model galaxy used to estimate the AMRDs of the UMi dSph.
			These SSPs are used to make the synthetic CMD by a linear combination with the weights.
			By comparing the synthetic CMD to observed CMD, we estimate the AMRDs.
			The detail of AMRD estimation is described in \secref{HGA}.
			The 391 SSPs correspond to $6\sim14$ Gyr with a $0.5$ Gyr interval, and $-3.2 \leq [\rm{Fe/H}] \leq -1.0$ with a $0.1$ dex interval.
			This age range is sufficient to solve the SFH found in fast dSphs\footnote{Fast dSphs lack extended star formation, with the majority of their stellar mass having been formed in a burst of star formation in the early universe.} (\cite{2015ApJ...811L..18G}) such as the UMi dSph, since star formation was quenched in the early universe.
			To make SSPs, we use the BaSTI isochrone, which is well suited to the properties of Galactic dwarf galaxies (e.g., wide range of $\alpha$-element abundances and metallicities).
			The $\alpha$-element abundance and the helium fraction are $[\rm{\alpha/Fe}] = 0.4$ and $Y = 0.247+1.31\Delta Z$, respectively.
			Dwarf galaxies typically show $\mathrm{[\alpha/Fe]} \sim 0.4$, though a weak negative trend is present over the metallicity range $-4 < \mathrm{[Fe/H]} < -1$ \citep{2004AJ....128.1177V}.
			Although $\mathrm{[\alpha/Fe]}=0.4$ is fixed considering the selection range of the BaSTI, the influence of the value of $\mathrm{[\alpha/Fe]}$ is judged to be negligible because there are few metal-rich stars exceeding $\mathrm{[Fe/H]}>-1.5$ in the UMi dSph.
			%To generate the CMD of SSPs, we create a stellar mass and magnitude function from the isochrone.
			We then adopt the stellar mass function as described in \secref{IMF} considering the effect of binaries.
			%The magnitude of binary systems is calculated as
			%\begin{equation}
			%	\label{eq:bin_mag}
			%	\mathrm{M_{bin}=M_{pri} - 2.5\times \log_{10}\left(1+\frac{10^{-0.4\times M_{comp}}}{10^{-0.4\times M_{pri}}}\right)},
			%\end{equation}
			%where, $\mathrm{{M_{bin}}}$ is the magnitude of binary, $\mathrm{M_{pri}}$ is the magnitude of primary star, and $\mathrm{M_{comp}}$ is the magnitude of companion star.
			In addition, we consider the photometric error of stars for the SSPs of each $g$- and $i$-band based on the observed data (see \figref{photerr}).
			Then, using these photometric errors, we add artificial errors assuming Gaussian forms to the magnitudes of each star of SSPs.
			We also convert the absolute magnitude to the apparent magnitude using the distance modulus of $(m-M)_{0}=19.18$ \citep{1999AJ....118..366M}.
			Since the observed data have been corrected for extinction and completeness effects (see \secref{reduction}), no further corrections are applied to the SSPs.
%			\indent
%			\figref{ssp_cmd} is the CMD made from the BaSTI isochrone such as the age of 12 $\rm{Gyr}$, the metallicity of $[\rm{Fe/H}]=-2.0$, the helium fraction $Y = 0.247+1.31\Delta Z$, and the alpha abundance of $\rm{[\alpha/Fe]}=-0.4$, and we assume the initial mass of the system, $10^{7}\ \rm{M_{\odot}}$, the binary fraction $f=0.5$, and mass ration of binary system is uniform distribution as indicated in \figref{massratio}. %			\begin{figure}[H]
%				\centering
%				\includegraphics[width=10cm]{./images/analysis/SSP_FEmH200_A120.png}
%				\caption{
%					SSP created from BaSTI, with age set to 12 Gyr, $\alpha$ abundance to $[\rm{\alpha/Fe}]=0.4$, metallicity to $[\rm{Fe/H}]=-2.0$, helium fraction to $Y=0.275$, binary fraction 0.5, and distant modulus $m-M=19.16$. 
%				\label{fig:ssp_cmd}}%labelはcaption内に書く必要あり
%			\end{figure}
		\subsection{Hybrid genetic algorithm}\label{HGA}
			We make the synthetic CMD by a linear combination with the weights of the SSPs. 
			These weights correspond to the star formation rate (SFR; \cite{1997AJ....114..680A}) of each population.
			We compare the observed CMDs with the synthetic CMDs in binned space.
			To find the best combination of weights for the synthetic CMD, we used the chi-square value \citep{1999ApJ...518..380M} defined as
%			\begin{eqnarray}
%				\chi^2 &=& \sum_{i=1}^{n_{\rm{bins}}} \frac{\big(\rm{{Count}_{\mathit{i}, obs}} + \min(\rm{{Count}_{\mathit{i}, obs}},1)}{\rm{{{Count}_{i, obs} + 1}}}\nonumber \\
%				&&- \frac{\rm{{Count}_{\mathit{i}, contami}}+ \sum_{j=1}^{n_{\rm{SSP}}} W_{j} \times \rm{{Count}_{\mathit{ij}, syn}}\big)^{2}}{\rm{{Count}_{\mathit{i}, obs}} + 1},
%				\label{eq:chi-square}
%			\end{eqnarray}
			\begin{equation}
				\chi^2 = \sum_{i=1}^{n_{\rm{bins}}} \frac{\big[\rm{{Count}_{\mathit{i}, obs}} + \min(\rm{{Count}_{\mathit{i}, obs}},1)-\rm{Count}_{\mathit{i}, model} \big]}{\rm{{Count}_{\mathit{i}, obs}} + 1}^{2},
				\label{eq:chi-square}
			\end{equation}
			where $\rm{{Count}_{\mathit{i}, obs}}$ indicates the number of stars in each bin of the 2D histogram of the observed CMD, $\rm{{Count}_{\mathit{i}, model}}$ is that of synthetic CMD, and ${n_{\rm{bins}}}$ is the number of bins in the CMD.
			To avoid division by zero and reduce the influence of the added 1 in the denominator, the term $\min(\rm{{Count}_{\mathit{i}, obs}},1)$ is added to the numerator as a matching correction \citep{1999ApJ...518..380M}.
			For $\rm{{Count}_{\mathit{i}, model}}$, we use
			\begin{equation}
				\rm{Count}_{\mathit{i}, model} = \rm{{Count}_{\mathit{i}, contami}}+ \sum_{j=1}^{n_{\rm{SSP}}} \mathit{W_{j}} \times \rm{{Count}_{\mathit{ij}, syn}},
			\end{equation}
			where ${n_{\rm{SSP}}}$ is the number of bins in the Hess diagram\footnote{In this paper, we refer to the Hess diagrams as the CMDs.}.
			We adopt a bin size of $0.075\times 0.025$ mag for $i_{0}$ magnitude and $(g-i)_{0}$ color, respectively.
			The $\rm{{Count}_{\mathit{i}, obs}}$ indicates the number of stars in each bin of the 2D histogram of the observed CMD.
			The $\rm{{Count}_{\mathit{ij}, syn}}$ is the number of stars in each bin of the 2D histogram of the CMD of each SSP.
			The $\rm{{Count}_{\mathit{i}, contami}}$ is the number of stars in each bin of the 2D histogram of the CMD of background and foreground contamination.
			$W_{j}$ is the weight of each SSP (corresponding to the SFR), thus $\rm{{Count}_{\mathit{i}, contami}}+\sum_{j=1}^{n_{\rm{SSP}}} \mathit{W}_{j} \times \rm{{Count}_{\mathit{ij}, syn}}$ represents the number of stars in each bin of the 2D histogram of synthetic CMD.

			These weights $W_{j}$ are interpreted as the SFRs for arbitrary SSPs with combinations of age and metallicity. 
			The distribution of these weights across age and metallicity space provides the reconstructed star formation history (SFH) and chemical evolution history (CEH). 
			%The resulting distributions are shown in the top-right panels of \figref{art_single}, right panels of \figref{art_amrd}, and panels of \figref{obs_amrd}.
			% We optimize the synthetic CMD with SSPs using the Hybrid Genetic algorithm (HGA; \cite{2015ApJ...811...76C}) that is the combination of Genetic Algorithm (GA) and the annealing.
			% %To recover the SFH and MDF of UMi dSph simultaneously, we compare the observed CMDs with the synthetic CMDs in pixelated space.
			% %To optimize the synthetic CMD to observed CMD, we adapt the Hybrid Genetic Algorithm (HGA) constructed by Cignoni et al. (2015): the Star Formation Evolution Recovery Algorithm (SFERA), and we also improve it to estimate the MDF.
			% %The star formation rates (SFR) of each population corresponding to the weight of the linear combination of 391 SSPs.
			% %HGA searches for the best fit synthetic CMD made from the linear combination of SSPs by comparing with the observed CMD using the residual of both. 
			% The GA models the mechanism of biological evolution and is suitable for searching for globally optimal solutions through processes such as crossover and mutation (e.g., \cite{1995ApJS..101..309C}). 
			% The annealing is a local search optimization technique that was inspired by metallurgical engineering (e.g., \cite{1983Sci...220..671K}). 
			% It accepts if solutions become worse, with a probability that depends on the number of trials. 
			% The synergy of GA and the annealing combines the advantages of both. 	
			We optimize the synthetic CMDs with SSPs using the Hybrid Genetic Algorithm (HGA; \cite{2015ApJ...811...76C}), which combines the strengths of the Genetic Algorithm (GA) and simulated annealing. 
			The GA, inspired by the principles of biological evolution, is well suited for global optimization through stochastic processes such as crossover and mutation (e.g., \cite{1995ApJS..101..309C}). 
			Simulated annealing, by contrast, is a local optimization method originally developed in metallurgical engineering (e.g., \cite{1983Sci...220..671K}), which allows the acceptance of less optimal solutions with a probability that gradually decreases during the search process.
			The hybridization of these two techniques leverages their complementary strengths: the GA efficiently explores a broad parameter space to avoid being trapped in local minima, while simulated annealing enhances convergence by fine-tuning solutions in the local vicinity. 
			This synergy enables a more robust and efficient search for the optimal star formation and chemical evolution histories that best reproduce the observed CMD, particularly in complex, multidimensional parameter spaces where traditional optimization methods may struggle.
		\subsection{Verification of the algorithm accuracy using artificial galaxies}\label{test}
			%m-Mのバリエーションでmetllicityがどのくらい変わるかも試す？？
			We evaluate the accuracy of peak estimation in the AMRD and the accuracy of our algorithm using artificial galaxies with a given AMRD.
			We perform two tests : the artificial galaxies with (1) a single population, and (2) multiple populations.
			The aim of the test (1) is to check that the peak is reasonably estimated in both metallicity and SFH over the range of metallicities.
			%to check that peak estimation in both metallicity and SFH is reasonably performed over the range of metallicities using in this study.
			The test (2) aims to verify that our algorithm detects the peaks of two populations with different metallicity and SFH in AMRD.
			%detects the separation on AMRD from the CMD of a dwarf galaxy with two populations with different metallicity and SFH.

\begin{figure}[h]
					\begin{center}
					\includegraphics[width=8.5cm]{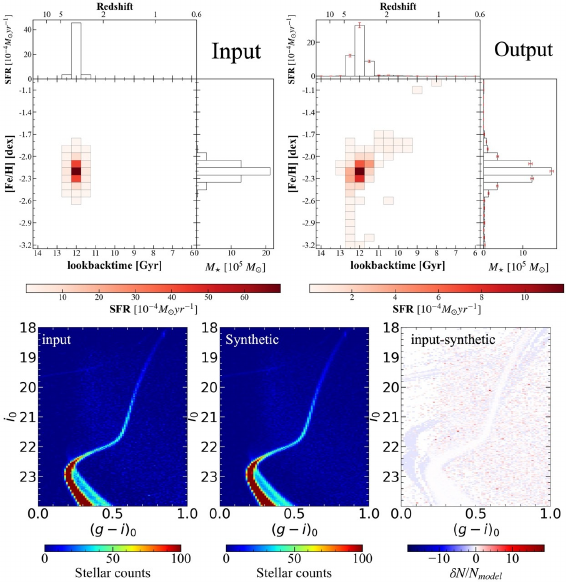}
					\end{center}
					\caption{
					Top left panel is the artificial AMRD assuming the single star burst with metallicity $\rm{[Fe/H]}=-2.2$ and $\rm{Age}=12\ \rm{Gyr}$ with metallicity dispersion of 0.01 dex and an age dispersion of 0.05 Gyr (Test (1)).
					Top right panel is the derived AMRD.
					In both of the top panels, the right histogram shows the metallicity distribution, and the top histogram shows the SFH. 
					These errors are estimate by the resampling-bootstrapping.
					Bottom left panel is the CMD of artificial galaxy.
					Bottom center panel is the derived synthetic CMD by our algorithm.
					Bottom right panel shows the difference fraction of both CMDs. 
					We adopt bins size of $0.075\times 0.025$ mag for $i_{0}$ magnitude and $(g-i)_{0}$ color, respectively.
					Colors represent the number counts of stars per bin.
					The $\delta N$ is the difference between input CMD and output CMD.
					The $N_{\rm{model}}$ is the stellar counts of the output.
					{Alt text: Five two-dimensional histograms.}
					% The bottom left panel is the CMD of artificial galaxy.
					% The bottom center panel is the derived synthetic CMD by our algorithm.
					% The bottom right panel shows the difference fraction of both CMDs.
					% We adapt bins size of $0.075\times 0.025$ mag for $i_{0}$ magnitude and $(g-i)_{0}$ color, respectively.
					% Colors represent the number counts of stars per bin.
					% The $\delta N$ is the difference between input CMD and output CMD.
					% In all these top panels, the right histogram shows the metallicity distribution, and the top histogram shows the SFH.
					% These errors are estimate by the resampling-bootstrapping.				
					\label{fig:art_single}
					}
				\end{figure}

\begin{figure}[h]
					\begin{center}
					\includegraphics[width=8.8cm]{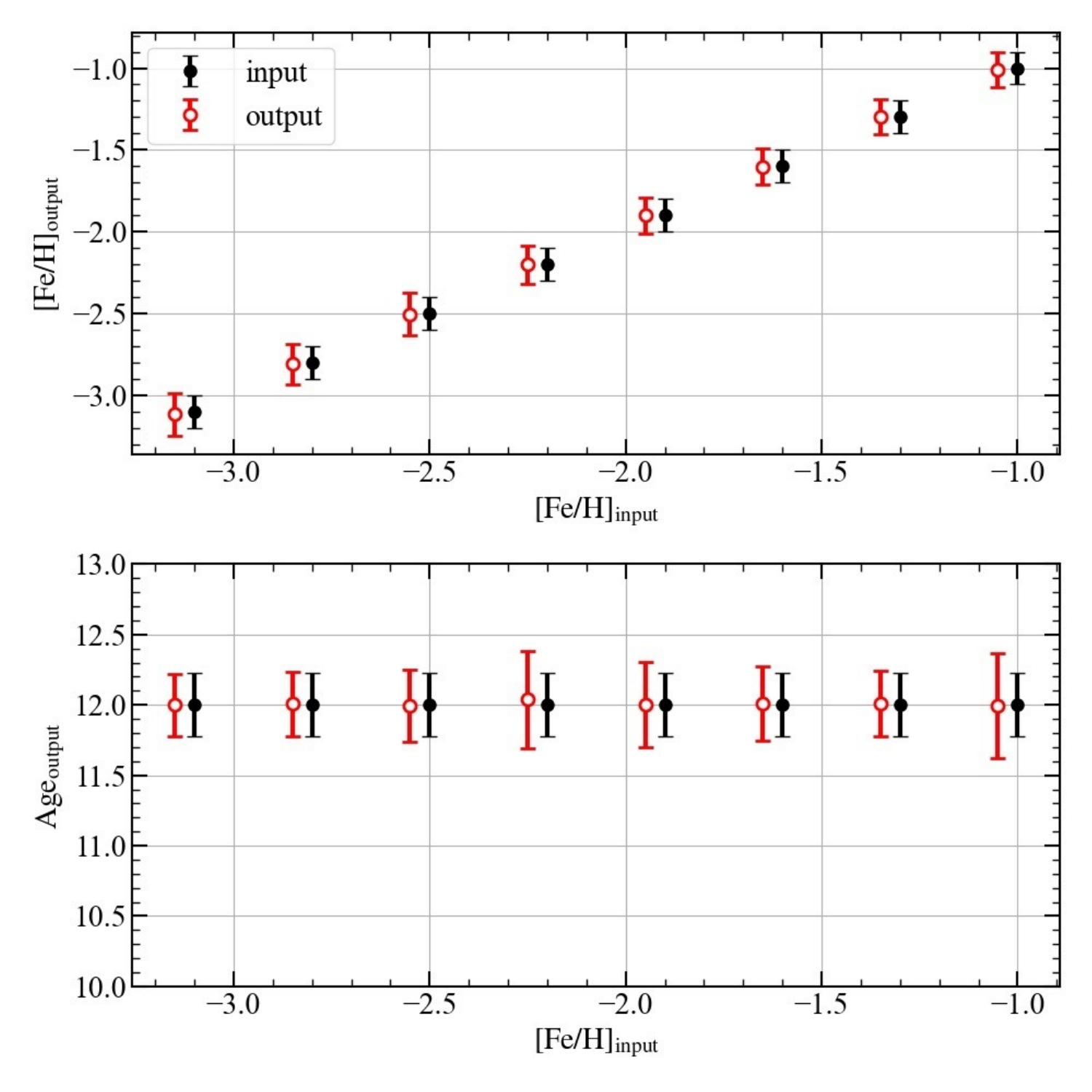}
					\end{center}
					\caption{Top panel shows the comparison of metallicity between input (black points) and estimated value (red open circles) of single-population artificial galaxies.
					We added a 0.05 offset for the input $[\rm{Fe/H}]$ on the x-axis for visibility.
					Bottom panel shows the comparison of the age of star formation peak between input and output from our algorithm.
					Error bars indicate the $1\sigma$ range, and the points show the mean value estimated by the Gaussian fit to the metallicity distribution.
					{Alt text: Two graphs.}
					\label{fig:evaluation}
					}
				\end{figure}

			For the test (1), we create CMDs of the artificial galaxy by combining SSPs with artificial AMRDs that account for the Gaussian SFH and metallicity distribution, and add the CMD of the contamination.
			The parameters for the artificial galaxy is given in \tabref{prm_artificial}.
			The contamination is taken from the rectangle area of ($240^{\circ} \leq \rm{R.A.} \leq 246^{\circ}, +42.5^{\circ} \leq\rm{Dec.} \leq +44.5^{\circ}$) from the Hyper Suprime-Cam Subaru Strategic Program DR3 \citep{2022PASJ...74..247A}.
			The selected area is located at a Galactic latitude of $b = 46.7^\circ$, similar to that of the Ursa Minor dSph ($b = 44.8^\circ$).  
			The Galactic longitudes of the selected field and Ursa Minor dSph are $l = 67.9^\circ$ and $l = 104.9^\circ$, respectively.  
			Although there is a difference in Galactic latitude, the field lies at $b \sim 45^\circ$, which is sufficiently far from the Galactic plane.  
			Therefore, the contamination from the disk is expected to be negligible in this case.
			%We retrieve the star catalog from the Wide survey of HSC/SSP DR3 \citep{2022PASJ...74..247A}.
			%To insert the contamination to CMD of artificial galaxy, then we carry out the resampling with duplication.
			We sample stellar objects from the contamination catalog with duplicates allowed, and add them to the CMD of the artificial galaxy.
			These CMDs are in the range of $18\leq i_{0} \leq 24$ and $0\leq (g-i)_{0} \leq 1.0$.
			The numbers of member stars and contamination are set to $80,000:20,000$ within the set CMD range.
			We prepare AMRDs that have a single population with $\rm{[Fe/H]} = -1.0$ to $-3.1$ at 0.3 dex interval, and the star formation peak is fixed at $\rm{Age} = 12\ \rm{Gyr}$.
			The standard deviation of the metallicity is fixed at 0.01 dex, and that of the star formation peak is 0.05 Gyr.
			The input and output of a case with $\rm{[Fe/H]} = -2.2$ and $\rm{Age} = 12\ \rm{Gyr}$ are shown in top panels of \figref{art_single}.
			The calculation parameters for the HGA are exhibited on \tabref{prm_obs}.
			We run the algorithm on these artificial CMDs with 100 times resampling bootstrapping and estimate the AMRDs.
			In addition, we fit the Gaussian function to the resulting SFHs and metallicity distributions.
			The comparison of the input SFHs and metallicity distributions value and the estimated value is shown in \figref{evaluation}.
			The error of output is estimated by variances in the 100 times of resampling bootstrapping.
			These error bars show the $1\sigma$ range of input and output.
			An example of derived AMRD and synthetic CMD are shown in the top right panel and bottom middle panel of \figref{art_single}.
			We confirm that our algorithm estimates the mean metallicity and star formation peak within the $1\sigma$ range of the inputs even at the lower metallicities such as $\rm{[Fe/H]} = -3.1$.
			However, in \figref{art_single}, the false sequence toward the metal-poor edge is detected.
				\begin{table*}[h]
					\renewcommand{\arraystretch}{1.5}
					\caption{Settings of artificial galaxies in test (1), test (2)-A and test (2)-B.}
					\label{table:prm_artificial}%info of obs
				\centering
					\footnotesize
					\begin{tabular}{cllllll}
						\hline
						Test code & $\rm{Mean}_{\rm{Age}}$ [Gyr] & $\rm{Mean}_{\rm{[Fe/H]}}$ [dex] & $\sigma^{2}_{\rm{Age}}$ [Gyr]& $\sigma^{2}_{\rm{[Fe/H]}}$ [dex]& $\rm{N_{\rm{member}}} : \rm{N}_{\rm{contamination}}$\\
						\hline
						Test (1) & $12$ & $-1.0$ to $-3.1$ with a $0.3$ dex interval  & $0.05$ & $0.01$ & $80,000:20,000$\\
						Test (2)-A & $12 $ and $ 12$ & $-2.0$ and $-2.3$ & $0.05 $ and $ 0.05$ & $0.01$ and $0.01$ & $80,000:20,000$\\
						Test (2)-B & $12$ and $13$ & $-2.0$ and $-2.3$ & $0.05 $ and $ 0.05$ & $0.01$ and $0.01$ & $80,000:20,000$\\
						\hline
					\end{tabular}
				\end{table*}

\begin{figure*}[t]
					\begin{center}
					\includegraphics[width=18cm]{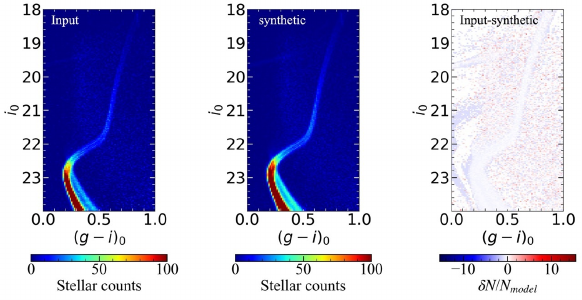}
					\end{center}
					\caption{
					The left panel shows the CMD of the artificial galaxy assuming simultaneous star formation 12 Gyr ago and $\rm{[Fe/H]} = -2.0, -2.3$ (Test (2)-A).
					These two populations have the same stellar mass. % relative to the total stellar mass of this artificial galaxy.
					The center panel shows the synthetic CMD derived by our algorithm.
					The right panel shows the difference fraction between observed and synthetic CMDs.
					The $\delta N$ is the difference between input and output CMD.
					The $N_{\rm{model}}$ is the stellar counts of outputs.
					{Alt text: Three two-dimensional histograms.}
					\label{fig:art_cmd}
					}
				\end{figure*}

\begin{figure*}[t]
					\begin{center}
					\includegraphics[width=15cm]{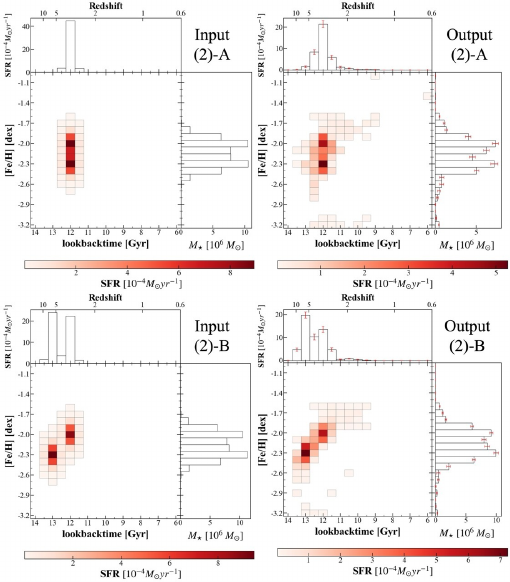}
					\end{center}
					\caption{
					The top two panels are the artificial and inferred AMRDs assuming simultaneous star formation for these two populations in Age = 12 Gyr (Test (2)-A).
					The bottom two panels assume different SFHs for them in Age = 12 and 13 Gyr (Test (2)-B).
					For both analyses these two population have different metallicity with $\rm{[Fe/H]} = -2.0, -2.3$.
					{Alt text: Four graphs.}
					\label{fig:art_amrd}
					}
				\end{figure*}

			For the test (2), we create two artificial galaxies that have two distinct populations of different metallicities $\rm{[Fe/H]}=-2.0, -2.3$.
			%In the case of (2)-A, we assume that both populations have a simultaneous star formation peak at 12 Gyr ago, and for (2)-B, that they have different age peaks at 12 and 13 Gyr ago.
			In test (2)-A, both populations have a simultaneous star formation peak at 12 Gyr ago. In test (2)-B, they have different star formation peaks at 12 and 13 Gyr ago.
			For each population we set as a 2-dimensional Gaussian distribution with an age dispersion of 0.05 Gyr and a metallicity dispersion of 0.01 dex.
			We set these two populations have a same number of stars ($40,000:40,000$).
			\begin{table}[h]
					\renewcommand{\arraystretch}{1.5}
					\caption{Calculation parameters for HGA with the model galaxies in test (1), test (2)-A and test (2)-B.}
					\label{table:prm_obs}%info of obs
				\centering
					\footnotesize
					\begin{tabular*}{8cm}{@{\extracolsep{\fill}}cc}
						\hline
						Parameter & Value \\
						\hline
					Number of generations & $5,000$\\
					Number of genes & $500$\\
					Mutation probability & $0.3$\\
					Number of digits of each weight & $4$\\
					Mutation number & $1,000$\\
					Initial temperature& $2,001$\\
					Cooling efficiency & $0.001$\\
					loops for simulated annealing& $200,000$\\
					resampling boot straps & $100$\\
						\hline
					\end{tabular*}
			\end{table}\\
			These artificial input AMRDs are shown on the left panels of \figref{art_amrd}.
			From these AMRDs we make the artificial CMDs.
			The CMD of the artificial galaxy with simultaneous star formation is shown in the left panel of \figref{art_cmd}.
			We apply our algorithm to these CMDs with the same calculation parameters used in test (1).
			The derived AMRDs are shown in the right panel of \figref{art_amrd}.
			As a result, our algorithm fully detects the two star formation peaks in both cases.
			In the (2)-A, the shape of the estimated metallicity distribution is almost identical to the input one.
			However, in the case of different SFHs, the ratio of metal-rich and metal-poor peaks in the histogram of metallicity distribution and SFH is different from the input AMRD.
			In particular, the young and metal-rich populations have a false detection of another population around high metal sequences ($\rm{[Fe/H]}\sim-1.7$ and $\rm{Age}\sim 10\ \rm{Gyr}$).
			This false detection is also visible in the case of simultaneous star formation.
			This false detection is likely attributable to the presence of contamination. 
			Notably, when the same estimation procedure was applied to a synthetic CMD constructed without adding contamination, no such false detection was observed.
			Also, the metal-poor population around $\rm{[Fe/H]} \sim -3.2$ is detected in both output (2)-A, and (2)-B in \figref{art_amrd}.
			These clumps are potentially contaminated in the AMRDs derived by the observational data referred to in the next \secref{AMRD_obs}.
		\subsection{Age metallicity relation diagram of the UMi dSph}\label{AMRD_obs}
%			Investigating whether stars of different metallicities in dwarf galaxies come from different progenitors is important for understanding how the first galaxies, such as UMi dSph, evolved in the early Universe.
%			It is important to examine whether stars with different metallicity populations originate in dwarf galaxies to understand the evolution of the candidates of first galaxies, such as UMi dSph, in the early Universe.
			To understand the evolution of candidate galaxies formed first in the early Universe, such as the UMi dSph, it is important to investigate how the stars having distinct metallicities formed in dwarf galaxies.
			Each stellar population identified on the AMRDs and its SFH will provide a clue to understanding the events (e.g., mergers, tidal interactions) experienced in the early epoch. 
			The AMRD allows us to obtain the detailed and complete SFH and accurate CEH (\cite{2012A&A...544A..73D}).
			Therefore, we estimate the AMRDs of the UMi dSph and their spatial difference using the HSC datasets and our algorithm described in \secref{HGA}. 

\begin{figure}[h!]
					\begin{center}
					\includegraphics[width=8.5cm]{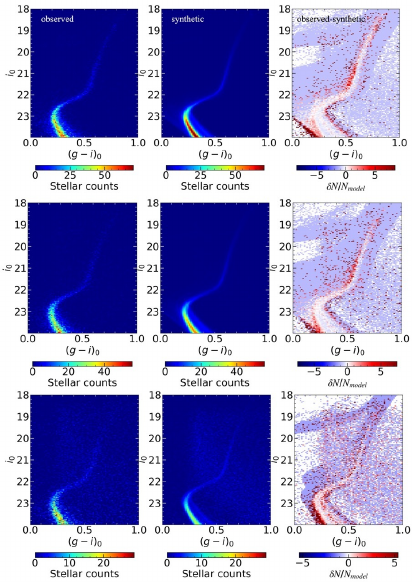}
					\end{center}
					\caption{
					The observed CMD (left), synthetic CMD derived by our algorithm (center), and difference fraction CMD (right) each of three annuli (Top: $r\leq r_{\rm{h}}$, Middle: $r_{\rm{h}}<r\leq2 r_{\rm{h}}$, Bottom: $2r_{\rm{h}}<r\leq6 r_{\rm{h}}$).
					The $\delta N$ is the difference between input and output CMD.
					The $N_{\rm{model}}$ is the stellar counts of outputs.
					{Alt text: Nine two-dimensional histograms.}
					\label{fig:cmd_obs}
					}
				\end{figure}

\begin{figure*}[t]
					\begin{center}
					\includegraphics[width=17cm]{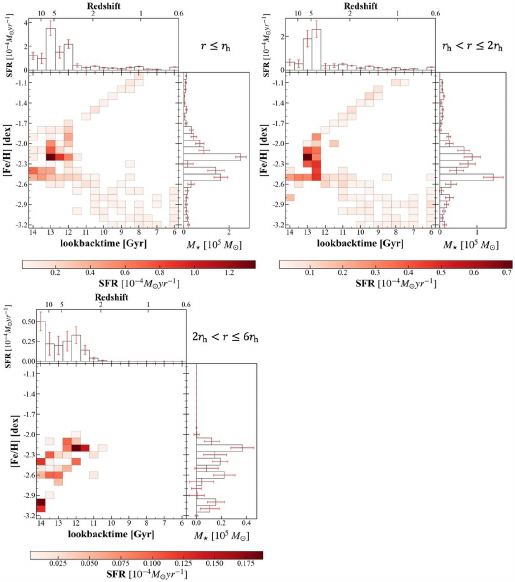}
					\end{center}
					\caption{
					The derived AMRDs from the observed data (Top left: $r \leq r_{\rm{h}}$, Top right:  $r_{\rm{h}}<r\leq2 r_{\rm{h}}$ and bottom left: $\ 2r_{\rm{h}}<r\leq6 r_{\rm{h}}$).
					the right histogram shows the metallicity distribution, and the top histogram shows the SFH. 
					These errors are estimate by the resampling-bootstrapping.
					{Alt text: Three two-dimensional histograms.}
					\label{fig:obs_amrd}
					}
				\end{figure*}

			The parameters for the HGA are set as the same setting with the \secref{test} (see \tabref{prm_artificial}).
			Regarding the contamination, we use the same control field used in \secref{test}.
			We separate the observed area into three regions depending on the distance defined with the elliptical distance ($r_{\rm{h}},\ 2 r_{\rm{h}},\ 6r_{\rm{h}}$, which correspond to the red dashed lines in \figref{CMD}) from the UMi center and obtain AMRDs for each region. 
			From the ratio of the area between these regions and the control field, we estimate the $\rm{Count}_{i,contami}$ of \equaref{chi-square}.
			In this calculation the $\chi^{2}$ values are converged.
				% \begin{table}[h]
				% 	\renewcommand{\arraystretch}{1.5}
				% 	\caption{Calculation parameters for HGA with observed data}
				% 	\label{table:prm_obs}%info of obs
				% \centering
				% 	\footnotesize
				% 	\begin{tabular*}{8cm}{@{\extracolsep{\fill}}cc}
				% 		\hline
				% 		Parameter & Value \\
				% 		\hline
				% 	Number of generations & $5000$\\
				% 	Number of gene & $500$\\
				% 	Mutation probability & $0.3$\\
				% 	Number of digits of each weight & $4$\\
				% 	Mutation number & $1000$\\
				% 	Initial temperature& $2001$\\
				% 	Cooling efficiency & $0.001$\\
				% 	loops for annealing& $200000$\\
				% 	resampling boot straps & $100$\\
				% 		\hline
				% 	\end{tabular*}
				% \end{table}\\
			\Figref{cmd_obs} shows the observed CMDs and synthetic CMDs and difference fraction of three annuli.
			The left panels are the observed CMDs that corrected the effect of reddening and detection completeness.
			In \figref{obs_amrd} the AMRDs in $r\leq r_{\rm{h}}$, $r_{\rm{h}}<r\leq2 r_{\rm{h}}$ and $2r_{\rm{h}}<r\leq6 r_{\rm{h}}$ of the UMi dSph are shown.
			All three regions have two distinct metallicity populations at $\rm{[Fe/H]}=-2.2$ and $-2.5$.
			These two distinct metallicity populations are consistent with the results of previous spectroscopic studies.
			\citet{2020MNRAS.495.3022P} identified two distinct chemodynamical populations within \timeform{30'}, which approximately corresponds to $2r_{\rm{h}}$ in our estimation.
			The first population has $\rm{[Fe/H]}=-2.05\pm0.03$ with $\sigma_{v}=4.9^{+0.8}_{-1.0}\ \rm{km\ s^{-1}}$.
			The second population has $\rm{[Fe/H]}=-2.29^{+0.05}_{-0.06}$ with $\sigma_{v}=11.5^{+0.9}_{-0.8}\ \rm{km\ s^{-1}}$.
			Our resulting two populations have peaks at $\rm{[Fe/H]}=-2.2$ and $\rm{[Fe/H]}=-2.5$, being comparable to those in \citet{2020MNRAS.495.3022P}.
			We find a metallicity offset of 0.2 dex between our results and those reported by \citet{2020MNRAS.495.3022P} for both metal-rich and poor populations.
			This offset may be influenced by uncertainties in the distance modulus.  
			To evaluate the effect of distance uncertainty on the RGB color, we compared the mean colors of isochrones at $[\rm{Fe/H}] = -2.2$ with distance modulus of 19.12 and 19.30, the latter of which reflects the uncertainty reported by \citet{1999AJ....118..366M}.  
			We find that the isochrone color $(g-i)_0$ becomes redder by 0.008 mag when the distance modulus increases.

			These metal-poor populations are $\approx$ 1 Gyr older than the metal-rich populations in all regions.
			In addition, the ratio of metal-rich and poor population changes with the distance from the center of the UMi dSph.
			According to the metallicity distribution indicated in \figref{obs_amrd}, the outer region is more dominated by metal-poor population than the inner region of $r\leq r_{\rm{h}}$.
			% The stellar mass ratio between metal-rich ($-2.4< \rm{[Fe/H]} \leq -2.0$) and metal-poor ($-2.6\leq \rm{[Fe/H]} \leq -2.4$) populations are ${M_{\odot,\rm{metal-rich}}}/{M_{\odot,\rm{metal-poor}}} = 1.51, 1.26, 1.29$ for $r \leq r_{\rm{h}},\ r_{\rm{h}}<r\leq2 r_{\rm{h}}$ and ,$\ 2r_{\rm{h}}<r\leq6 r_{\rm{h}}$, respectively.
			% This result is in agreement with the metallicity gradient found in the UMi dSph \citep{2022A&A...665A..92T}.
			% However, this stellar mass ratios are not consistent with the result of \citet{2020MNRAS.495.3022P}, ${M_{\odot,\rm{metal-rich}}}/{M_{\odot,\rm{metal-poor}}} =1.15$
			% By comparing the spatial distribution of \citet{2020MNRAS.495.3022P} to our HSC data, the data of \citet{2020MNRAS.495.3022P} tend to distribute along the major axis.
			The stellar mass ratios between the metal-rich ($-2.4 < \rm{[Fe/H]} \leq -2.0$) and metal-poor ($-2.6 \leq \rm{[Fe/H]} \leq -2.4$) populations are ${M_{\odot,\rm{metal-rich}}}/{M_{\odot,\rm{metal-poor}}} = 1.51$, $1.26$, and $1.29$ for the radial bins $r \leq r_{\rm{h}}$, $r_{\rm{h}} < r \leq 2r_{\rm{h}}$, and $2r_{\rm{h}} < r \leq 6r_{\rm{h}}$, respectively.
			This result is consistent with the previously reported metallicity gradient in the UMi dSph \citep{2022A&A...665A..92T}. 
			However, our stellar mass ratios are not in agreement with the result of \citet{2020MNRAS.495.3022P}, reported a lower ratio of $1.15$.
			A comparison of the spatial distribution of stars in \citet{2020MNRAS.495.3022P} with our data suggests that their sample is more concentrated along the major axis of UMi. 
			Due to the smaller sample size and the relative lack of stars along the minor axis in their dataset, it is possible that their result is affected by sample selection bias.

			%In the case of $\ 2r_{\rm{h}}<r\leq6 r_{\rm{h}}$ region, this ratio become higher than inner region.
			%We suspect that the reason for this is that this result does not include metal-poor stars than $\rm{[Fe/H]} <-2.6$
			Several formation mechanisms to form these multiple chemodynamical populations have been proposed; e.g., mergers (\cite{2019MNRAS.488.2312G}, \cite{2015MNRAS.454.3996D}), stellar feedback \citep{2016ApJ...820..131E, 2021MNRAS.501.5121M}, interaction with cosmic filaments \citep{2019MNRAS.488.2312G}, and tidal interaction with the host galaxy during the infall of a dwarf galaxy \citep{2019MNRAS.488.2312G}.
			Our resulting AMRDs are similar to the simulated AMRDs of figure 9 in \citet{2019MNRAS.488.2312G} that assumes a past merged dwarf galaxy in the early Universe.
			However, this scenario does not fully account for the 1 Gyr offset observed between the SFH of the metal-rich and metal-poor populations.
			Further discussion of the evolution history of the UMi dSph is described in \secref{comparison}.

			We also find that the diagonally distributed young and metal-rich population is visible around age $\sim 9$ Gyr and $[\rm{Fe/H}]\sim-1.4$.
			This population is visible in the inner two regions ($\leq 2r_{\rm{h}}$).
			This metal-rich population has already been discovered in the central region of the UMi dSph by the spectroscopic survey \citep{2010ApJS..191..352K}.
			%This population could formed in the remnants gas of UV-feedback or stellar feedback
			%ここではSN feedbackとかionizationで残ったガスから化学濃縮された話をしたい。
			%特にUV-feedback Becker, R. H., et al. 2001, AJ, 122, 2850 で ionization はz=6くらいと予想されていて、わたしたちの結果の星形成の転換期と一致している。
			According to \citet{2019MNRAS.488.2312G}, investigating the CEH of a post-merger dwarf galaxy in a cosmological simulation, these metal-rich populations formed by the accreting gas ejected by stellar feedback.
			%There are two theories that these stars have survived the process of reionization due to the effect of self shielding (\cite{2004ApJ...600....1S}; \cite{2011ApJ...730...14H}; \cite{2010MNRAS.402.1599S}), and that they were formed by accreted gas after star formation was quenched by stellar feedback. 
			%It is conceivable that these stars are reservoirs of self-shielding gas and have survived the process of re-ionization (\cite{2004ApJ...600....1S}; \cite{2011ApJ...730...14H}; \cite{2010MNRAS.402.1599S}).

			In addition, the fake feature is visible in \figref{obs_amrd}.
			The metal-poor and young ($[\rm{Fe/H}]\lesssim-2.6$ and $\rm{age}\lesssim11$ Gyr) distributions are visible except the most outer region ($2 r_{\rm{h}}\leq r$).
			This feature is a false detection of the blue straggler as young stars.
			In the synthetic CMD of \figref{cmd_obs}, we see the stellar components having similar color and magnitude with those of the blue straggler around $(g-i)_0\sim0.1$ and $i_0\sim23$.
			In our algorithm, the SSPs used to fit the observed CMD do not reproduce blue straggler stars since the BaSTI isochrone does not calculate their evolutionary tracks.			Therefore, we conclude that this population is not a genuine young metal-rich population, but the blue stragglers misclassified as the young populations in our algorithm.
			
			The errors in this analysis are composed of the random errors and the systematic errors.
			The random errors are estimated by the 100 times resampling bootstrapping.
			Then the results of the $\ 2r_{\rm{h}}<r\leq6 r_{\rm{h}}$ region has a larger error than other two regions, since it contains fewer number stars and the contamination is dominant.
			The systematic error depends on the isochrone.
			To estimate this effect, we use Dartmouth isochrones \citep{2008ApJS..178...89D} instead of BaSTI for our algorithm and compare the results in the \appref{apd:dartmouth}.
			In summary, the systematic error by adopting the BaSTI isochrones would not be significant.

	\section{Discussion}\label{discussion}
		\subsection{The multiple populations in the color distribution of RGB}\label{comparison}
			To obtain the supporting evidence on the bimodal metallicity distribution we detected in \secref{AMRD_obs}, we investigate the color distribution of RGB stars within the $r\leq r_{\rm{h}}$ region.
			In \secref{AMRD_obs}, the metallicity estimation is influenced not only by RGB stars but also by MSTO and MS stars, as the entire CMD is used for the fitting.
			Therefore, we test whether the bimodal feature is still visible in the RGBs color distribution, which is sensitive to metallicity.
			Note that the color of RGBs also depends on stellar age: at fixed metallicity, younger stars appear bluer on the RGB.
			However, the color shift due to age is generally smaller than that caused by differences in metallicity.
			In the following analysis, we test for the presence of a bimodal color distribution in the RGB, taking into account not only the effect of metallicity but also the color variation due to age.

\begin{figure*}[t]
				\begin{center}
				\includegraphics[width=18cm]{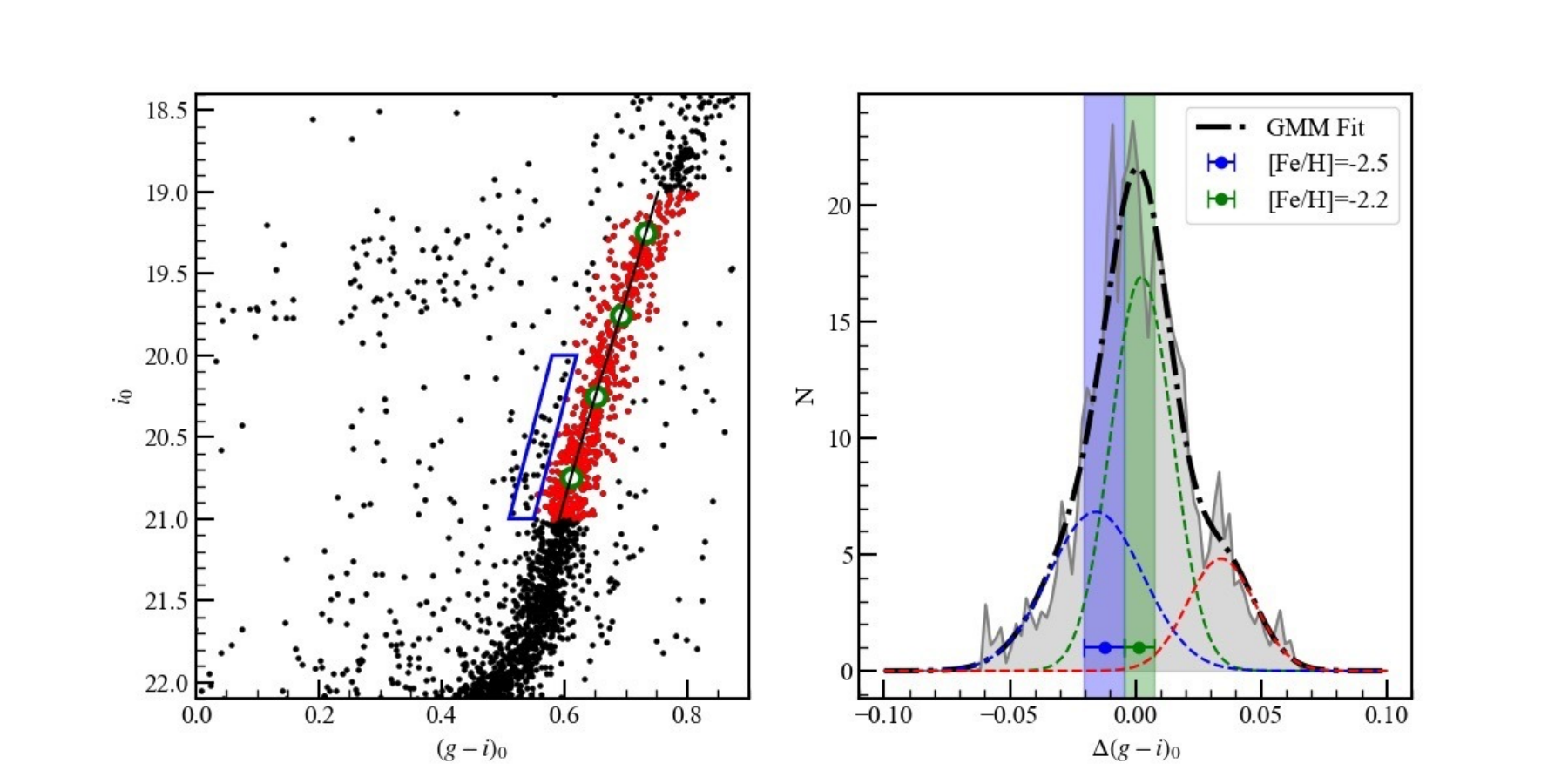}
				\end{center}
				\caption{
					The left panel is the CMD of the $r \leq r_{\rm{h}}$ region of the UMi dSph. 
					The red points are the selected RGB stars. 
					The open green points are the mean $(g - i)_0$ values of each subgroup. 
					The black solid line is the result of fitting with the cubic spline. 
					The blue parallelogram covers the distribution of binary RGBs.
					The right panel shows the color distribution of the corrected stars to the $\Delta(g - i)_0$ color system and the result of the GMM fitting. 
					The shaded area is the kernel density distribution of each star considering the photometric error values of each star. 
					The colored points and shaded regions represent the color ranges of the corrected isochrones, corresponding to ages and metallicities of $12\ \rm{Gyr}$ with $\mathrm{[Fe/H]}= -2.2$ and $13\ \rm{Gyr}$ with $\mathrm{[Fe/H]} = -2.5$, in the $\Delta(g - i)_0$ color system.
					The black dash-dotted line show the fitted GMM and these blue, green and red plots represent the individual components.
					{Alt text: Two graphs.}
					% The left panel is the CMD of the $r\leq r_{\rm{h}}$ region of the UMi dSph.
					% The red points are the selected RGB stars.
					% The open green points are the mean $(g-i)_{0}$ values of each subgroup.
					% The black solid line is the result of fitting with the cubic spline.
					% The right panel shows the color distribution of the corrected stars to the $\Delta (g-i)_{0}$ color system and the result of the GMM fitting.
					% The shaded area is the kernel density distribution of each star considering the photometric error values of each star.
					% The colored points and shaded regions indicate the color range of the corrected isochrones with an age of 12 Gyr to the $\Delta (g-i)_{0}$ color system.
					% The black dash-dotted line show the fitted GMM and these blue, green and red plots represent the individual components.
					% The blue parallelogram covers the distribution of binary RGBs.
				\label{fig:CMD_bimo}
				}
			\end{figure*}

			We bin the stars in the range of $19 < i_{0} < 21$ into subgroups in 0.5 mag increments.
			%The stars in the range of $19 < i_{0} < 21$ are binned into subgroups in 0.5 mag increments.
			%The green open circles in the \figref{CMD_bimo} indicate the mean color and the center magnitude of each bin
			In this method, we find the blue stars next to the RGB sequence surrounded by the blue parallelogram.
			These stars could be binary RGBs, which are considered in the SSPs.
			Therefore, the effects of these blue RGBs are negligible in the derived AMRDs in \secref{AMRD_obs}.
			%These stars distributed uniformly, and assuming they are truly the blue RGB stars, they are more metal-poor than $\rm{[Fe/H]}=-3.2$ (the limit of BaSTI).
			%We fit the young and metal-poor isochrones to this sequence, then well mached.
			For all subgroups, we limit the relative color $(g-i)_{0}$ within $\pm0.06$ for the benchmark isochrone ($\rm{[Fe/H]}=-2.2$, $\rm{Age}=12 \rm{Gyr}$). 
			The threshold ($\pm0.06$) is adopted by eye inspection in order to exclude possible binary RGBs, foreground and background contamination, and horizontal branch stars.
%			For all subgroups, the $(g-i)_{0}$ threshold is set to $\pm0.06$ by eye in order to avoid including this binary RGBs, contamination, and HB stars.
			%These stars possibly be yellow straggler stars mainly found in the MW open cluster and globular cluster.
			A single Gaussian is fit to the color distribution of each subgroup for estimating the mean color of RGBs at the magnitude. 
			The fitted standard deviations are all found to be smaller than 0.03 mag. 
			These mean color values and bins center magnitudes of subgroups are indicated in the left panel of \figref{CMD_bimo}, as open green circles.
			A cubic spline fitting is performed on all the means to obtain the reference RGB track for the $\Delta(g-i)_{0}$ color distribution.
			We calculated $\Delta(g-i)_{0}$ relative color of each star to the reference track.
			We constructed a color distribution of the RGB stars by stacking each star as a Gaussian with a standard distribution of a photometric error of the star as show in right panel of \figref{CMD_bimo}.
			Then, we perform the Gaussian Mixture Model (GMM) fit to the relative color distribution $\Delta(g-i)_{0}$.
			In the GMM fitting, the number of Gaussian components $n$ are conditioned on 1 to 4.
			We select the best model using Akaike information criterion (AIC; \cite{Akaike1998}).
			We define the difference in the AIC values relative to a minimum AIC value as $\Delta \rm{AIC}(n)$, where $n$ is the number of components in the GMM.
			The minimum $\rm{AIC}$ value is obtained with $n=3$ ($\Delta \rm{AIC} (1)=6.2$, $\Delta \rm{AIC} (2)=13.8$, and $\Delta \rm{AIC} (4)=2.3$).
			From this result, we consider that the color distribution of RGBs has three components.
			We also convert the $\rm[Fe/H]=-2.2$ with $12\ \rm{Gyr}$ and $\rm[Fe/H]=-2.5$ with $13\ \rm{Gyr}$ isochrones, which were detected as peaks in AMRD (see \figref{obs_amrd}) to the $\Delta(g-i)_{0}$ relative color.
			The peak values of two GMM components are located in the 1 $\sigma$ range of the color of the model isochrones.
			The number ratio between these components is $N_{\rm metal-rich} / N_{\rm metal-poor} = 1.62$.
			The mass ratio within the $r \leq r_{\rm h}$ region, estimated from the AMRD, is ${M_{\rm metal-rich}} / {M_{\rm metal-poor}} = 1.51$.
			These values are comparable.
			We presume that the reddest component at $\Delta (g-i)_0 \sim +0.035$ could be stars formed during the secondary star formation episode identified in AMRD.
			This population may have originated from gas accreted after being ejected by stellar feedback in $r \leq 2r_{\rm{h}}$ region, or it could be due to foreground contamination.
			In summary, the distribution of the relative color of RGB stars supports the multimodal metallicity distribution suggested by AMRD.

		\subsection{Photometric vs. Spectroscopic metallicity distributions}\label{mdf_comp}
\begin{figure}[h!]
			\begin{center}
			\includegraphics[width=8.5cm]{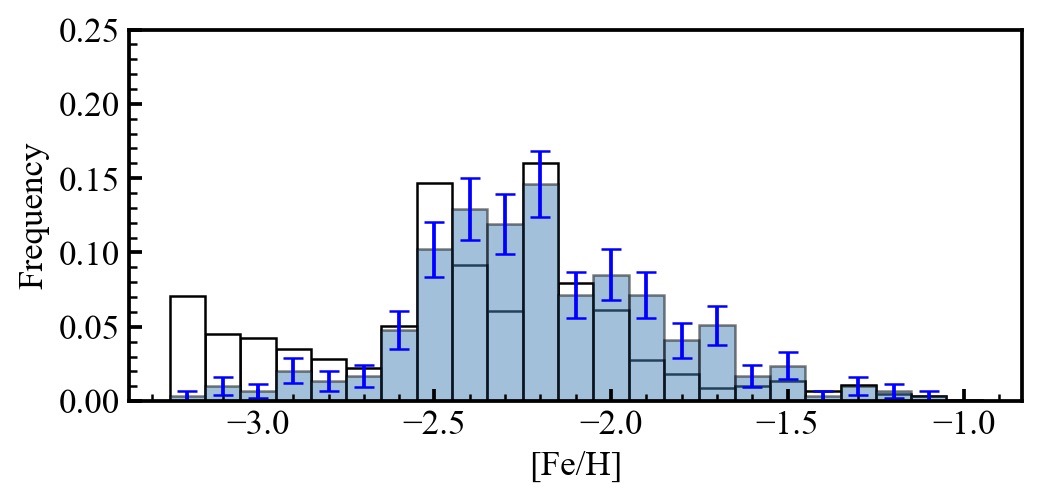}
			\end{center}
			\caption{
				Comparison of the metallicity distribution derived by our algorithms (black histogram) within $2 r_{\rm{h}}$ and of stars with reliable metallicity distribution (blue histogram) from \citet{2011ApJ...727...78K, 2013ApJ...779..102K, 2020MNRAS.495.3022P}.
				The spectroscopic sample is selected by the $\delta\mathrm{[Fe/H]} < 0.2\ \mathrm{dex}$.
				Red and blue error bars represent Poisson errors for the photometric and spectroscopic MDFs, respectively.
				{Alt text: A histogram.}
			\label{fig:mdf_comp}
			}
		\end{figure}

		To evaluate the consistency of our photometrically derived metallicity distribution with previous spectroscopic measurements, we compare our results with the spectroscopic metallicity distributions which constructed by the Keck/DEIMOS spectroscopic data \citep{2010ApJS..191..352K,2011ApJ...727...78K,2013ApJ...779..102K,2020MNRAS.495.3022P}.
		Considering the metallicity estimation error of our algorithm ($\delta\mathrm{[Fe/H]} \approx 0.2\ \mathrm{dex}$; see \figref{evaluation}), we select the spectroscopic data with $\delta\mathrm{[Fe/H]} < 0.2\ \mathrm{dex}$.
		%The photometric metallicity distribution in \figref{mdf_comp} is constructed by integrating the AMRDs along the age axis within two radial bins: $r \leq r_{\rm{h}}$ and $r_{\rm{h}} < r \leq 2 r_{\rm{h}}$, as described in \secref{AMRD_obs}. 
		The photometric metallicity distribution in \figref{mdf_comp} is constructed from RGB stars ($i_0 < 22$) in the synthetic CMDs. %within two radial bins: $r \leq r_{\rm{h}}$ and $r_{\rm{h}} < r \leq 2 r_{\rm{h}}$.
		This magnitude limit is set to be consistent with the spectroscopic data.
		Also, our photometric results are limited to stars within $2 r_{\rm{h}}$ in order to match the spatial coverage of the spectroscopic sample.
		% For the synthetic CMD, among the SSPs with different ages and metallicities that contribute to the RGB, we ignore the age information and use only the metallicity to derive the metallicity distribution.
		Our photometric results are limited to stars within $2 r_{\rm{h}}$ in order to match the spatial coverage of the spectroscopic sample.

		\Figref{mdf_comp} shows both the photometric and spectroscopic metallicity distributions.
		Both distributions display bimodal peaks.
		While the two distributions share the similar overall trend, the photometric metallicity distribution exhibits a more pronounced bimodal structure than the spectroscopic one.
		% We attribute this difference partly to the binning scheme used in our analysis, in which metallicities were discretized into 0.1 dex intervals according to the isochrones.
		% In this scheme, each star is assigned the metallicity of its best-fitting isochrone, rather than treating metallicity as a continuous parameter.
		% If a more flexible estimation approach were adopted, for example, by evaluating a continuous metallicity likelihood based on each star proximity to isochrones, the resulting metallicity distribution might differ in shape and.

		To improve the accuracy of photometric metallicity estimation in our algorithm, particularly at the low-metallicity end, we plan to incorporate metallicity-sensitive narrowband photometry (e.g., \cite{2017MNRAS.471.2587S}), such as NB395 which is installed in Subaru/HSC. 
		HSC/NB395 sensitive to features such as the Ca II H \& K lines \citep{1985AJ.....90.2089B} have proven effective in improving metallicity estimates, particularly at low metallicities, in combination with broadband photometry \citep{2017MNRAS.471.2587S}.
		\subsection{Spatial and metallicity-dependent SFH variations in the UMi dSph}\label{comparison}
\begin{figure*}[h!]
				\begin{center}
				\includegraphics[width=18cm]{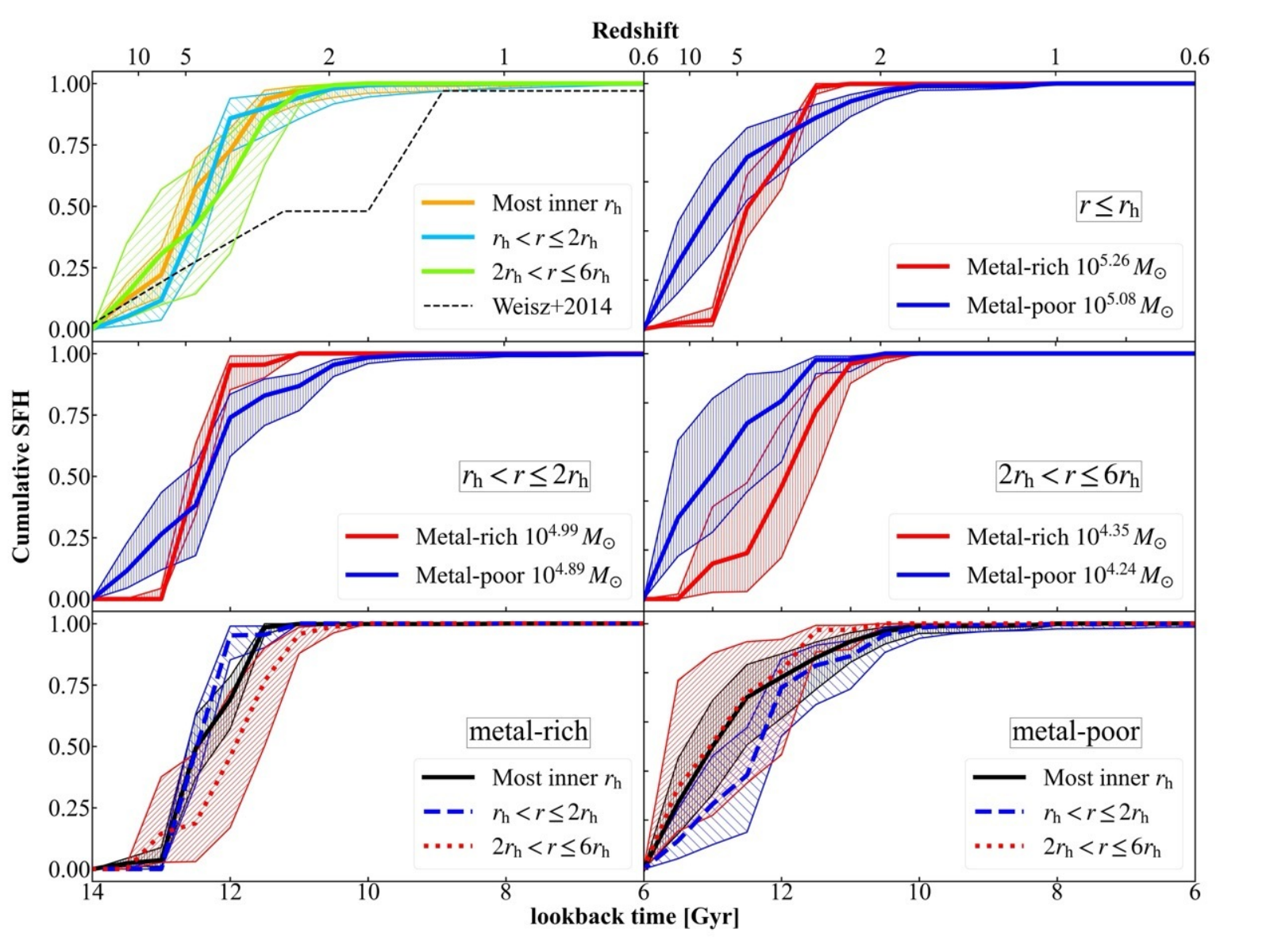}
				\end{center}
				\caption{
					Top left: the star formation histories of three different regions of the UMi dSph, depending on the distance from the center.
					The light blue line shows the cumulative SFH of the inner $r\leq r_{\rm{h}}$ region.
					The orange solid line is for $r_{\rm{h}}<r\leq2 r_{\rm{h}}$, and the green one is for $2r_{\rm{h}}<r\leq6 r_{\rm{h}}$.
					The dashed black line is the SFH of \citet{2014ApJ...789..147W}.
					Top-right and middle panels show the SFH of different metallicity populations in the different regions.									   
					The red solid line covers the range of $-2.4< \rm{[Fe/H]} \leq -2.0$.
					The blue solid line covers the range of $-2.6\leq \rm{[Fe/H]} \leq -2.4$.
					The total mass of each population is given in the legends. 
					The bottom two panels show the SFH of metal-rich and metal-poor populations comparing three different regions.
					{Alt text: Six graphs indicate the cumulative star formation histories.}
				\label{fig:sfh}
				}
			\end{figure*}

			The spatial and metallicity gradients of dwarf galaxies preserve a record of their formation history (e.g., \cite{2006A&A...459..423B}, \cite{2015MNRAS.454.3996D}, \cite{2017MNRAS.467..208O}). 
			The major merger in dSphs, for instance, drives older stars outward while younger stars are formed in the center, resulting in an "outside-in" age gradient \citep{2016MNRAS.456.1185B}.			
			Stellar feedback also induces metallicity gradients by dispersing metal-poor stars outward and fueling metal-rich star formation in the galaxy's center \citep{2021MNRAS.501.5121M}.  
			These processes lead to spatially different distributions of stellar populations with different age and metallicity. 
			In the UMi dSph, the metal-poor population exhibits a more extended and circular distribution ($\epsilon=0.33^{+0.12}{-0.09}$, $r_{\rm{h}}=286^{+13}{-12}\ \rm{pc}$), whereas the metal-rich population is more centrally concentrated and shows a more flattened distribution ($\epsilon=0.75\pm0.03$, $r_{\rm{h}}=221\pm17\ \rm{pc}$; \cite{2020MNRAS.495.3022P}).
			The origin of these distinct stellar populations is still not well understood. 
			Determining the SFHs of the metal-rich and metal-poor populations separately is therefore essential for understanding the physical processes that gave rise to these differences. 
			Investigating the SFH dependency on the distance from the center and metallicity will allow us to reveal how such different stellar populations have formed over time, and to uncover the formation and evolution history of the UMi dSph.
			Because our dataset covers more than six times the half-light radius of the UMi dSph, we are able to derive the SFH at different radial distances from the galaxy center. 
			This wide coverage allows us to investigate how both the SFH and the CEH vary with position, as discussed in \secref{AMRD_obs}.
			The SFHs are obtained by integrating the observed AMRDs along the metallicity axis, as shown in \figref{obs_amrd}. 
			The top left panel of \figref{sfh} presents the cumulative SFHs for three radial bins: the inner region ($r \leq r_{\rm{h}}$), an intermediate annulus ($r_{\rm{h}} < r \leq 2r_{\rm{h}}$), and the outer annulus ($2r_{\rm{h}} < r \leq 6r_{\rm{h}}$). 
			We restrict the metallicity range to $-2.6 \leq \mathrm{[Fe/H]} \leq -2.0$ to avoid contamination from misclassified populations and the secondary, younger metal-rich population that appears as a diagonal feature in \figref{obs_amrd}.

			As a reference, we overlay the SFH derived by \citet{2014ApJ...789..147W} using deep HST/ACS data for the central region of the UMi dSph. 
			Their study shows a pause of star formation between approximately 10 and 11 Gyr ago. 
			In contrast, our results do not show such a pause phase in any of the annuli, suggesting that star formation proceeded in a shorter timescale than previously reported.

			We further investigate the SFHs of two distinct metallicity populations: a metal-poor population ($-2.6 \leq \mathrm{[Fe/H]} \leq -2.4$) and a metal-rich population ($-2.4 < \mathrm{[Fe/H]} \leq -2.0$). 
			For each population, we compare the SFHs across the same three radial regions (inner, intermediate, and outer) to explore how their star formation histories depend on both metallicity and distance from the center of the UMi dSph.
			For the metal-poor population, stars in the intermediate annulus are slightly younger than those in the inner region, a trend that is consistent with previous findings by \citet{2002AJ....123.3199C}. 
			However, the SFH in the outer annulus is similar to that in the inner region, suggesting no clear age gradient across the full extent of the metal-poor population.
			In contrast, the metal-rich population exhibits a different pattern. 
			Star formation in the inner region continues for a longer duration than in the intermediate annulus. 
			Interestingly, in the outer annulus, star formation appears to continue even longer than in the inner two regions. 
			Despite these spatial variations in SFH, no systematic age gradient is observed within the metal-rich population either.
			
			Overall, while the SFHs of both metallicity populations vary with distance from the center, neither population exhibits a monotonic age gradient. 
			However, across all regions, the metal-rich population consistently forms about 1 Gyr later than the metal-poor population. 
			% These findings suggest a complex formation history for the UMi dSph, possibly involving a recent merger between dwarf galaxies with distinct stellar populations.
			These findings point to a complex formation history for the UMi dSph, potentially driven by a combination of dynamical and baryonic processes, including a recent merger between chemically distinct dwarf galaxies, and environmental interactions.

			%Especially, supporting this recent merger scenario, \citet{2020MNRAS.495.3022P} reported a signature of prolate rotation in spectroscopic data from the Multiple Mirror Telescope/Hectochelle, which may indicate a recent merger event \citep{2014MNRAS.445L...6L}. 
			Especially, \citet{2020MNRAS.495.3022P} reported a signature of prolate rotation in spectroscopic data from the Multiple Mirror Telescope/Hectochelle, which may indicate a recent merger event \citep{2014MNRAS.445L...6L}.
			In such a scenario, the observed spatial variations in SFH could be explained by the different spatial distributions of stars originally belonging to the progenitor systems.
			The recent merger may have caused these distinct stellar population to become spatially mixed, resulting in the spatially complex SFH we observe in the UMi dSph today.

			% Also, environmental interactions, such as pericentric passages and tidal effects, could account for the spatially complex SFH.
			% A pericentric passage could have triggered central star formation \citep{2019MNRAS.488.2312G}.
			% If the metal-rich population formed during such a pericentric passage and any pre-existing age gradient was subsequently disrupted by tidal interactions, the spatially complex SFH observed in the UMi dSph could be naturally explained.
			% According to \cite{2020ApJ...905..109M,2019arXiv190604180F}, the timing of first infall of the UMi dSph is expected in $\approx 11.0\ Gyr$ ago.
			% The peak of SFR of the metal-rich population around 12--13Gyr ago.
			% Considering the uncertainties of infall time (1.75 Gyr; \citet{2020ApJ...905..109M}), these timing is may be related.
			Environmental processes may also have played a role in shaping the spatially complex SFH.
			In particular, pericentric passages around the Milky Way could have triggered episodes of star formation in the center of dwarf galaxies \citep{2019MNRAS.488.2312G}, while subsequent tidal interactions might have disrupted any pre-existing age gradients.
			For the UMi dSph, the timing of its first infall is estimated to be $\approx 11.0$ Gyr ago \citep{2020ApJ...905..109M, 2019arXiv190604180F}, with an uncertainty of 1.75 Gyr\citep{2020ApJ...905..109M}.
			The peak in SFR of the metal-rich population around 12--13 Gyr ago could therefore plausibly be linked to environmental influences during early orbital passage, when the galaxy may have been near pericenter.

			Stellar feedback and secondary starbursts has been proposed as a mechanism for forming multiple stellar populations and spatial gradients.
			Repeated gas outflows and potential fluctuations could drive the radial migration of older stars \citep{2016ApJ...820..131E, 2019MNRAS.490.1186G}.
			However, the absence of a clear monotonic age gradient in either metallicity population of the UMi dSph implies that stellar feedback alone cannot account for the observed spatial distribution of stellar ages.
			It is also possible that stars formed during a secondary starburst were dynamically mixed by subsequent environmental effects, such as tidal interactions with the MW, resulting in the spatially complex SFH currently observed in the UMi dSph.
			% Similarly, in the case of the Sextans dSph, which is known to host multiple stellar populations with distinct metallicities and velocity dispersions, probable member stars that are more metal-rich than its mean metallicity have been found in its outer regions \citep{2025arXiv250402787T}.
			% According to \cite{2025arXiv250402787T}, if these stars are realistic member, the metallicity gradient which is commonly discovered in MW dwarf galaxies is eliminated. 
			% Taken together, these findings from both Sextans dSph and the UMi dSph suggest that the canonical picture of inside-out formation and metallicity gradient in dwarf galaxies may not be universal. 

			To further investigate these hypotheses, a spectroscopic survey of the UMi dSph with Subaru/Prime Focus Spectrograph (PFS) is currently in progress. 
			Chemodynamical analysis of this dataset will be crucial in uncovering the origins of multiple stellar populations in the UMi dSph. 
			If the galaxy underwent a recent merger, signatures such as residual rotation may still be observable, as the system may not yet have reached dynamical equilibrium.
\section{Summary}\label{summary}
	In this study, we have presented the structure and stellar populations of the UMi dSph. 
	Based on the wide-field photometric dataset obtained by Subaru/HSC, we constructed the deep CMD that reached below its MSTO and covered beyond the tidal radius of the UMi dSph. 
	This allows us to derive the detailed structural properties, which show a more circular shape than the previous estimation with the limited central area by \citet{2018ApJ...860...66M}

	In order to investigate the formation history of the UMi dSph, we have developed the algorithm to solve the SFH and CEH from the CMDs by combining the genetic algorithm and the annealing.
	We tested the accuracy of our algorithm using the CMDs of artificial galaxies.
	We created SSPs for each metallicity and applied our algorithm to obtain AMRDs.
	We confirmed that both the estimated metallicity distribution and SFH agreed with the input within 1 $\sigma$ at any metallicity.
	We also performed a separation test for the two distinct metallicity populations, assuming simultaneous SF and 1 Gyr difference cases.
	We confirmed that this algorithm is able to estimate AMRDs without any deviation in the position of each peak in both cases.

	We applied this algorithm to three regions divided according to the structural parameters described above, $r\leq r_{\rm{h}}$, $r_{\rm{h}}<r\leq2 r_{\rm{h}}$, and $2r_{\rm{h}}<r\leq6 r_{\rm{h}}$ to derive the SFH and CEH.
	We identified the two distinct metallicity populations reported in previous studies.
	The SFH of both metal-rich and metal-poor populations show spatial variations, without any age-gradients.
	The star formation of the  metal-rich population had started about 1Gyr later than that of the metal-poor one.
	The origin of this delay and spatially different formation histories may be related to a recent merger.

	In this work we detect complex SFHs depending on the spatial distribution and metallicity even in the UMi dSph, which is among the least massive of the classical dSphs and one of the fast dwarf.
	We will apply this algorithm to other dwarf galaxies with different stellar masses.
	For example, UFDs are a good targets to investigate weather smaller systems have a similar SFH to that of classical dSphs.
%%%%%%%%%%
\section*{Acknowledgments}\label{acc}
This work was supported by JSPS KAKENHI Grant Numbers 18H05875, 20K04031, 20H05855.
This work was supported by JST SPRING, Japan Grant Number JPMJSP2104.
RFGW acknowledges support through the generosity of Eric and Wendy Schmidt, by recommendation of the Schmidt Futures program.

The Hyper Suprime-Cam (HSC) collaboration includes the astronomical communities of Japan and Taiwan, and Princeton University. 
The HSC instrumentation and software were developed by the National Astronomical Observatory of Japan (NAOJ), the Kavli Institute for the Physics and Mathematics of the Universe (Kavli IPMU), the University of Tokyo, the High Energy Accelerator Research Organization (KEK), the
Academia Sinica Institute for Astronomy and Astrophysics in Taiwan (ASIAA), and Princeton University. 
Data analysis was carried out on the large-scale data analysis system co-operated by the Astronomy Data Center (ADC) and Subaru Telescope, NAOJ. 
This research is based on data collected at the Subaru Telescope.
We are honored and grateful for the opportunity of observing the Universe from Maunakea, which has the cultural, historical, and natural significance in Hawaii.
The data are obtained from SMOKA, which is operated by the NAOJ/ADC.

The Pan-STARRS1 Surveys (PS1) have been made possible through contributions of the Institute for Astronomy, the University of Hawaii, the Pan-STARRS Project Office, the Max-Planck Society and its participating institutes, the Max Planck Institute for Astronomy, Heidelberg and the Max Planck Institute for Extraterrestrial Physics, Garching, The Johns Hopkins University, Durham University, the University of Edinburgh, Queen's University Belfast, the Harvard-Smithsonian Center for Astrophysics, the Las Cumbres Observatory Global Telescope Network Incorporated, the National Central University of Taiwan, the Space Telescope Science Institute, the National Aeronautics and Space Administration under Grant No. NNX08AR22G issued through the Planetary Science Division of the NASA Science Mission Directorate, the National Science Foundation under Grant No. AST-1238877, the University of Maryland, and Eotvos Lorand University (ELTE).

The authors would like to thank the referee for constructive comments, which helped improve the quality of this paper.

\appendix
\section{MCMC result of the structural properties}\label{apd:cornerplot}
\begin{figure*}[h]
			\begin{center}
			\includegraphics[width=14cm]{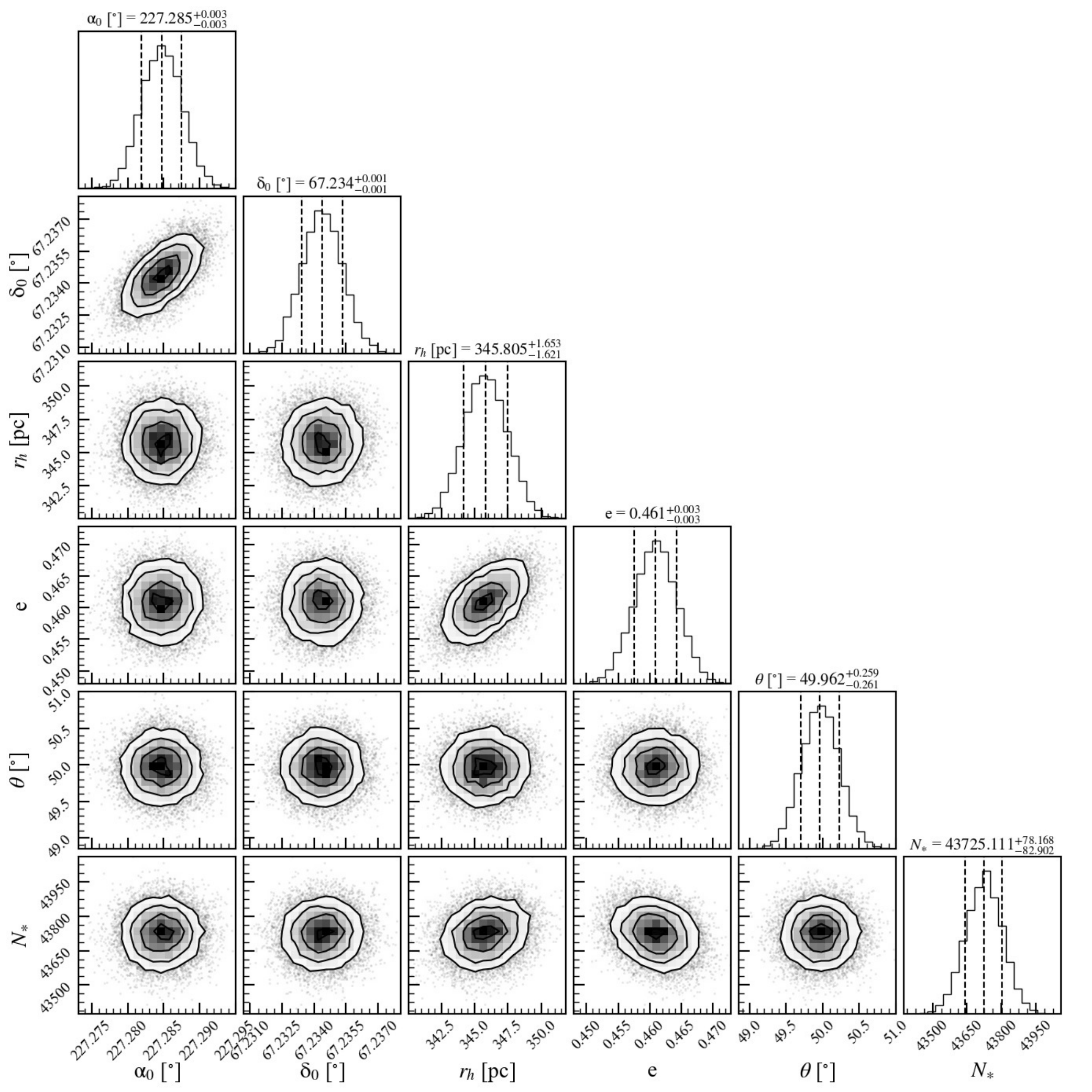}
			\end{center}
			\caption{
			The top panels of each column are marginalized posterior distribution for all estimated parameters: the central coordinate of the UMi dSph in the equatorial coordinates, $\rm{\alpha_{0}}$ and $\rm{\delta_{0}}$, the half-light radius, $r_{\rm{h}}$, the ellipticity, $\epsilon$, the position angle of the major axis, $\theta$, and the number of member stars, $N_{\ast}$. 
			The other panels are 2D probability density functions for the parameters.
			The dashed line indicates the median value of each parameter and 68\% Bayesian credible intervals. 
			The contours correspond to $1\sigma, 2\sigma$, and $3\sigma$ confidence intervals assuming Gaussian distributions.
			{Alt text: Fifteen contour plots and six histograms.}
			\label{fig:corner}
			}
		\end{figure*}

		In \figref{corner}, we show the MCMC result as corner plots (posterior distribution and marginal distributions for all parameters) in the case of the structural parameter estimation (see \secref{structuralProperties}).
\section{AMRDs with Dartmouth isochrone}
\label{apd:dartmouth}
		We also estimate the AMRDs of two inner regions, $r\leq r_{\rm{h}}$ and $r_{\rm{h}}<r\leq2 r_{\rm{h}}$, of the UMi dSph with the Dartmouth isochrone \citep{2008ApJS..178...89D}.
		The metallicity coverage in the metal-poor side of the Dartmouth isochrone is more limited than that of the BaSTI isochrone.
		To estimate the AMRDs in the Dartmouth isochrone, we prepare the 255 SSPs that correspond to $6--14$ Gyr with a $0.5$ Gyr interval, and $-2.45 \leq [\rm{Fe/H}] \leq 1.05$ with a $0.1$ dex interval.
		The $[\rm{\alpha/Fe}]$ is set as 0.4, and the helium fraction is set as $Y=0.245+1.5 \rm{Z}$.
		For the HGA calculation, we set the same parameters as in \tabref{prm_obs}.

\begin{figure*}[t]
			\begin{center}
			\includegraphics[width=18cm]{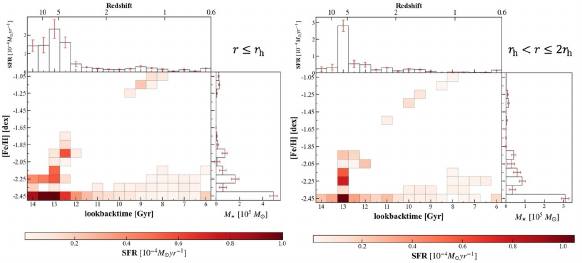}
			\end{center}
			\caption{
			The derived AMRDs of divided three regions  with the Dartmouth isochrones (Top left: $r \leq r_{\rm{h}}$, Top right:  $r_{\rm{h}}<r\leq2 r_{\rm{h}}$ and bottom left: $\ 2r_{\rm{h}}<r\leq6 r_{\rm{h}}$).
			Note that the metallicity coverage of the Dartmouth isochrones is limited on the metal-poor side, which may cause the metal-poor population to accumulate at the lowest metallicity bins.
			{Alt text: Two two-dimensional histograms.}
			\label{fig:dar_amrd}
			}
		\end{figure*}

		Results are shown in \figref{dar_amrd}.
		We confirmed the bimodal trend at $[\rm{Fe/H}]=-2.25$ and $[\rm{Fe/H}]=-2.45$ in both two regions.
		The total mass ratios of the metal-poor ($[\rm{Fe/H}]=-2.45$) population to the metal-rich ($-2.35\leq[\rm{Fe/H}]\leq-1.75$) population for the first 2.5 Gyr are 0.86 and 0.98 for $r\leq r_{\rm{h}}$ region and $r_{\rm{h}}<r\leq2 r_{\rm{h}}$ annulus, respectively.
		Due to the limited coverage of metallicity, the metal-poor population is overestimated.
		In order to explore the multiple populations of MW dSphs, it is essential to have a metallicity range that covers the metal-poor side so that it can adequately distinguish the two populations.
\bibliographystyle{apj.bst}
\bibliography{ref_umi}%*3

\end{document}